\documentclass[superscriptaddress,aalso receivedpaper,10pt,twocolumn,floats,showpacs,amsmath,amsfonts,amssymb,reprint,nofootinbib]{revtex4-2}

\usepackage{graphicx}
\usepackage[colorlinks=true,citecolor=blue]{hyperref}
\usepackage[caption=false]{subfig}
\usepackage{pstricks}
\usepackage{pst-node}
\usepackage{amsmath}
\usepackage{amsbsy}
\usepackage{amsfonts}
\usepackage{mathtools}
\usepackage{verbatim}
\usepackage{dsfont}
\usepackage{MnSymbol}
\usepackage{graphicx,color}
\usepackage[table]{xcolor}
\usepackage{enumitem}
\usepackage{bigints}

\newcommand{\re}{\text{Re}}

\newcommand{\Tr}{\text{Tr}}
\DeclareMathOperator*{\argmin}{arg\,min}

\begin{document}

	\title{Quantum instanton approach to metastable collective spin systems: \\ Numerical implementation and applications}

\author{Maciej Chudak}
 	\affiliation{Institute of Molecular Physics, Polish Academy of Sciences, Mariana Smoluchowskiego 17, 60-179 Pozna\'{n}, Poland}
    
	\author{Massimiliano Esposito}
\affiliation{Complex Systems and Statistical Mechanics, Department of Physics and Materials Science, University of Luxembourg, 30 Avenue des Hauts-Fourneaux, L-4362 Esch-sur-Alzette, Luxembourg}

\author{Krzysztof Ptaszy\'{n}ski}
	\email{krzysztof.ptaszynski@ifmpan.edu.pl}
 	\affiliation{Institute of Molecular Physics, Polish Academy of Sciences, Mariana Smoluchowskiego 17, 60-179 Pozna\'{n}, Poland}
	
	\date{\today}

\begin{abstract}
Driven-dissipative collective spin systems can host several long-lived metastable states; rare switching between them governs slow relaxation and first-order dissipative phase transitions associated with changes in the dominant stationary state. In the companion Letter [arXiv:2604.15091], we introduced a quantum instanton framework based on exact quasiprobability dynamics to describe this switching and illustrated it with a minimal model possessing a simplifying symmetry. Here we develop its numerical implementation, combining collocation and shooting methods, and apply it to two microscopically motivated systems: the Lipkin--Meshkov--Glick model with purely collective dissipation and the cooperative resonance fluorescence with collective and local dissipation. By determining the leading exponential scaling of switching rates between metastable attractors, we predict the stationary magnetization, the first-order dissipative phase-transition point, and the exponential scaling of slow relaxation rates. These predictions agree with quantum master-equation calculations and an exact analytical result for the second model without dephasing. We also examine how the choice of quasiprobability representation affects numerical convergence and find that the alternative semiclassical Wigner approach, based on Gaussian truncation, has varying accuracy across models and parameter regimes. These results substantiate the framework's applicability to experimentally relevant collective spin systems with both collective and local dissipation.

\end{abstract}

\maketitle

\section{Introduction}
Driven-dissipative quantum systems can have several long-lived metastable states even when their finite-size dynamics has a unique stationary state~\cite{macieszczak2021theory,carr2013nonequilibrium,rodriguez2017probing,fink2018signatures,chen2023quantum,beaulieu2025observation,noachtar2026exact}. This situation commonly arises when their mean-field dynamics develops multiple attractors in a large-system limit. At finite size, fluctuations induce rare transitions between their basins of attraction~\cite{drummond1989quantum,kinsler1991quantum,lee2012colllective}, as observed, for example, in nonlinear quantum resonators~\cite{rodriguez2017probing,fink2018signatures,chen2023quantum,beaulieu2025observation}. These transitions determine both the long relaxation time and the relative stationary occupations of the metastable states. When a control parameter changes the dominant (most likely) attractor, the corresponding crossover sharpens with system size and transforms into a first-order dissipative phase transition in the thermodynamic limit~\cite{minganti2018spectral,chen2023quantum,beaulieu2025observation,roberts2023exact}.

Mean-field stability alone does not determine which attractor dominates: the relevant information is contained in the exponentially small switching rates between metastable states. For attractors $i$ and $j$, their leading dependence on a large parameter $V$ takes the Arrhenius-like form~\cite{dykman1988quantum,dykman2007critical,drummond1989quantum,kinsler1991quantum,lee2024real,xiang2025switching}
\begin{align} \label{eq:arrh}
\kappa_{i\rightarrow j} \asymp \exp(-V\mathcal{A}_{i \rightarrow j}) \,.
\end{align}
Here $\mathcal{A}_{i\rightarrow j}$ is \emph{activation barrier}, and $\asymp$ denotes equality at the leading exponential order, $a \asymp b \Leftrightarrow \lim_{V \rightarrow \infty} (\ln a)/(\ln b)=1$. In a two-attractor reduction, the stationary occupations satisfy $P_i/P_j=\kappa_{j\rightarrow i}/\kappa_{i\rightarrow j}$; hence their relative weight is set, to exponential accuracy, by the difference $\mathcal{A}_{i\rightarrow j}-\mathcal{A}_{j\rightarrow i}$. A crossing of the barriers can thus locate a first-order transition, while their magnitudes govern the exponentially slow switching and relaxation.

Collective spin systems offer an important setting for this problem, with realizations and proposals in cavity-based atomic systems~\cite{mivehvar2021cavity,morrison2008dynamicalprl,morrison2008collective,norcia2018cavity,muniz2020exploring,song2025dissipation,ferri2021emerging}  and solid-state platforms~\cite{mlynek2012demonstrating,nissen2013collective,kessler2012dissipative,shakirov2016role,ferreira2019lipkin}. We use this term for ensembles of $N$ spins whose dynamics is invariant under permutation of the spins, and is thus naturally expressed through collective spin operators $\hat J_k=\sum_{i=1}^{N}\sigma_k^i/2$ ($k=x,y,z$), where $\sigma_{x,y,z}^i$ denotes the local Pauli operator of the $i$th spin. They can be further divided into systems with \emph{strong} and \emph{weak} permutation symmetry. In the strong case, the Hamiltonian and each jump operator can be written in terms of collective spin operators and thus commute with $\hat{\mathbf J}^{2}=\sum_k\hat J_k^2$. The total-spin quantum number $J$, defined by the eigenvalue $J(J+1)$ of $\hat{\mathbf J}^{2}$, is then conserved. Within a fixed-$J$ sector, the dynamics admits a macrospin description, and $J$ is the large parameter $V$. Under weak permutation symmetry, the Liouvillian remains invariant under spin permutation, but independent, identical local dissipative channels can mix  sectors with different $J$~\cite{shammah2018open}. This situation matters for cavity spin ensembles with both collective and local dissipative channels~\cite{song2025dissipation}; in this case, the large parameter is $V=N$.

The companion Letter~\cite{ptaszynski2026quantum} introduced a  framework for calculating activation barriers $\mathcal{A}_{i\rightarrow j}$ in such settings from exact, untruncated equations of motion for quantum quasiprobability distributions. In the large-system limit, these equations yield effective dynamics governed by an auxiliary classical Hamiltonian; the actions of its rare switching trajectories, called \emph{instantons}, give the barriers. Using a minimal model selected for its simplifying symmetry, the Letter showed how these barriers predict both the dominant stationary state and the exponential scaling of the relaxation rate. It also demonstrated that the semiclassical Wigner approach~\cite{dubois2021semi,gronchi1979fokker,dutta2025quantum,xiang2025switching}, by truncating higher-order derivatives in the quasiprobability dynamics and thereby discarding non-Gaussian contributions, can shift the predicted transition and alter the relaxation-rate scaling. Unlike auxiliary Hamiltonians associated with classical Markov processes, those derived from quantum quasiprobability dynamics can be nonconvex in momentum. This precludes the direct use of standard minimum-action algorithms~\cite{grafke2019numerical,gagrani2023action,zakine2023minimum} and makes the numerical construction and selection of physically relevant instantons a central challenge.

Here we develop a numerical implementation of the instanton framework that combines collocation and shooting methods, and apply it to two microscopically motivated systems with different mechanisms of metastability. In the driven-dissipative Lipkin--Meshkov--Glick (LMG) model with strong permutation symmetry, metastability arises due to competition of Hamiltonian spin interactions and collective decay~\cite{morrison2008dynamicalprl,morrison2008collective,wang2021dissipative}. Cooperative resonance fluorescence (CRF), relevant to a driven atomic ensemble in an optical cavity~\cite{song2025dissipation}, combines coherent driving and collective emission with local decay and dephasing; it has weak permutation symmetry. These examples test the instanton framework under distinct sources of nonlinearity and dissipation without relying on the full set of symmetries that simplifies the Letter's model.

We use the computed barriers to predict steady-state magnetization, first-order transition boundaries, and the exponential scaling of relaxation rates, and compare these predictions with quantum master equation calculations. For the CRF model without dephasing, an exact solution supplies a further benchmark~\cite{roberts2023exact,noachtar2026exact}; the finite dephasing case tests the approach beyond that solvable limit. We also examine how the nonconvexity of the auxiliary Hamiltonians and non-Gaussian effects neglected by the semiclassical Wigner approach manifest in the studied models. 

Our paper is organized as follows. Section~\ref{sec:framework} recalls the instanton framework and introduces its numerical implementation; Secs.~\ref{sec:lmg} and~\ref{sec:crf} present the two applications, followed by conclusions in Sec.~\ref{sec:concl}. The Appendixes provide supporting derivations, computational details, and further discussion.

\section{Instanton framework} \label{sec:framework}

\subsection{Theory outline}
The instanton approach developed in the companion Letter~\cite{ptaszynski2026quantum} is constructed from the dynamics of an appropriate quasiprobability distribution $p_\alpha$, 
\begin{align} \label{eq:evquas}
\partial_t p_\alpha (\boldsymbol{x},t)=L_\alpha(\boldsymbol{x},\nabla_{\boldsymbol{x}}) p_\alpha(\boldsymbol{x},t) \,,
\end{align}
where $L_\alpha(\boldsymbol{x},\nabla_{\boldsymbol{x}})$ is the differential operator generating the dynamics, $\boldsymbol{x}$ is the vector of the system coordinates, and $\nabla_{\boldsymbol{x}}$ is the gradient over $\boldsymbol{x}$. The coordinate vector $\boldsymbol{x}$ may correspond, e.g., to coordinates on the Bloch sphere for systems with a strong permutation symmetry or magnetization components in systems with weak permutation symmetry. Specific representations $\alpha$ are discussed for the examples in Secs.~\ref{sec:lmg} and~\ref{sec:crf}.

Specifically, this approach is applicable to systems in which the differential operator $L_\alpha$ obeys an appropriate scaling of its derivative terms with the macroscopic parameter $V$, namely,
\begin{align}
L_\alpha(\boldsymbol{x},\nabla_{\boldsymbol{x}})=\sum_{\boldsymbol{n}} V^{1-\Vert \boldsymbol{n}\Vert_1} \partial^{\boldsymbol{n}}_{\boldsymbol{x}} a_{\alpha,\boldsymbol{n}}(\boldsymbol{x}) \,,
\end{align}
where $\boldsymbol{n}=(\ldots,n_k,\ldots)$ are vectors of nonnegative integers with $\Vert\boldsymbol{n}\Vert_1=\sum_{k} n_k \geq 1$, and $\partial^{\boldsymbol{n}}_{\boldsymbol{x}} \equiv \prod_k \partial_{x_k}^{n_k}$. Then, the large-$V$ asymptotic behavior of metastable switching is determined by an auxiliary Hamiltonian
\begin{align} \label{eq:auxhamdef}
\mathcal{H}_\alpha(\boldsymbol{x},\boldsymbol{\pi})=\lim_{V \rightarrow \infty} V^{-1} L_\alpha(\boldsymbol{x},-V \boldsymbol{\pi}) \,,
\end{align}
where we replace each derivative $\partial_{x_k}$ in the differential operator with a scalar $-V \pi_k$. This Hamiltonian generates the corresponding
Hamilton's equations,
\begin{align} \label{eq:hamilton-eq}
d_t \boldsymbol{x}=\nabla_{\boldsymbol{\pi}} \mathcal{H}_\alpha (\boldsymbol{x},\boldsymbol{\pi}) \,, \quad d_t \boldsymbol{\pi}=-\nabla_{\boldsymbol{x}} \mathcal{H}_\alpha (\boldsymbol{x},\boldsymbol{\pi}) \,.
\end{align}

The dynamics $d_t \boldsymbol{x}$ for $\boldsymbol{\pi}=\boldsymbol{0}$ is equivalent to mean-field (MF) dynamics of the system collective variables, which becomes exact in the limit $V \to \infty$. In the multistable regimes considered here, mean-field dynamics can have several long-time solutions, called attractors, depending on initial conditions. In particular, our paper focuses on a situation when the attractors are stable fixed points (FPs) $\boldsymbol{x}^*$ for which $d_t \boldsymbol{x}=\boldsymbol{0} \, \vert_{\boldsymbol{x}=\boldsymbol{x}^*}$; we leave the case of time-dependent attractors (e.g., limit cycles) for future consideration. The stability of the fixed points is determined by the Jacobian
\begin{align} \label{eq:mfjacobian}
\mathbb{J} \equiv \left[\partial_{x_k}\dot{x}_l \right]_{kl}\, \vert_{\boldsymbol{x}=\boldsymbol{x}^*} \,.
\end{align}
The fixed point is then stable if all Jacobian eigenvalues have negative real parts, while it is unstable when at least one of them is positive\footnote{Throughout this article, we omit the issue of nonhyperbolic fixed points with zero Jacobian eigenvalues, as they are confined to a measure-zero subset of parameter space.}.

As mentioned in the Introduction, for large but finite $V$, the mean-field attractors correspond to long-lived metastable states of the system. As shown in the companion Letter~\cite{ptaszynski2026quantum}, the activation barriers $\mathcal{A}_{i \to j}$ governing the stochastic switching between them [see Eq.~\eqref{eq:arrh}] can be calculated as the action along an instanton trajectory satisfying Hamilton's equations~\eqref{eq:hamilton-eq},
\begin{align}
\mathcal{A}_{i \to j}= \lim_{T \to \infty} \int_0^T dt  \boldsymbol{\pi}(t) \cdot d_{t} \boldsymbol{x}(t) \,,
\end{align}
where $\boldsymbol{\xi}(t)=[\boldsymbol{x}(t),\boldsymbol{\pi}(t)]$
is the instanton trajectory with the boundary conditions 
\begin{align} \label{eq:boundcondinst}
\boldsymbol{\xi}(0)=(\boldsymbol{x}_i^*,\boldsymbol{0}) \,, \quad \boldsymbol{\xi}(T)=(\boldsymbol{x}_j^*,\boldsymbol{0}) \,, \quad T \to \infty \,,
\end{align}
where $\boldsymbol{x}_{i,j}^*$ is the coordinate vector of the fixed point $i,j$. When several trajectories satisfy these boundary conditions, the one with the lowest action determines the activation barrier. This selection among candidate trajectories is distinct from minimizing the action over all paths satisfying $\boldsymbol{x}(0)=\boldsymbol{x}_i^*$ and $\boldsymbol{x}(T)=\boldsymbol{x}_j^*$, as in classical stochastic systems with convex auxiliary Hamiltonians (see Sec.~\ref{subsec:lmgnonconvex}). Specifically, when the separatrix between the basins of attraction of fixed points $i$ and $j$ contains one or more unstable FPs $\nu$, the instanton trajectory is composed of two segments: 
\begin{enumerate}
    \item $\boldsymbol{\pi}\neq \boldsymbol{0}$ segment associated with positive action, connecting stable FP $i$ to one of the unstable FPs $\nu$;
     \item $\boldsymbol{\pi}= \boldsymbol{0}$ segment associated with zero action, corresponding to mean-field relaxation trajectory from the unstable FP $\nu$ to the stable FP $j$.
\end{enumerate}
Thus, because the relaxation segment contributes no action, the calculation of the activation barrier reduces to finding the $\boldsymbol{\pi}\neq\boldsymbol{0}$ segment of the instanton.

\subsection{Numerical implementation} \label{subsec:framework-num}

The instanton satisfying the conditions above is an infinite-time trajectory. The simplifying symmetry of the model in the companion Letter~\cite{ptaszynski2026quantum} allowed us to construct it exactly. That construction does not extend to the more complex models studied here, which require a numerical approximation. For classical stochastic systems with auxiliary Hamiltonians convex in $\boldsymbol{\pi}$, efficient minimum-action algorithms provide such an approximation~\cite{grafke2019numerical,gagrani2023action,zakine2023minimum}. The auxiliary Hamiltonians considered here are instead nonconvex in $\boldsymbol{\pi}$ (see Secs.~\ref{subsec:lmgnonconvex} and~\ref{subsubsec:crfnonconvex}), so these algorithms cannot be applied. We therefore use two boundary-value methods to approximate the $\boldsymbol{\pi}\neq\boldsymbol{0}$ instanton segment by a finite-time trajectory with boundary conditions approximating the exact asymptotic ones.

\subsubsection{Collocation method} \label{subsubsec:collocation}

The first method solves a two-point boundary-value problem by collocation: it approximates the trajectory within a finite-dimensional space of trial functions and enforces Hamilton's equations~\eqref{eq:hamilton-eq} at selected points, together with the boundary conditions specified below. We implement it using SciPy~\cite{2020SciPy-NMeth} and its \texttt{solve\_bvp} routine~\cite{kierzenka2001bvp,ascher1995numerical}.

For a $d$-dimensional state vector $\boldsymbol{x}$, the $2d$ unknown initial phase-space coordinates $\boldsymbol{x}(0), \boldsymbol{\pi}(0)$ and the unknown duration $T$ require $2d+1$ scalar boundary constraints. After testing several formulations, we adopted the following set. The first $2d$ conditions constrain the initial point of the $\boldsymbol{\pi}\neq\boldsymbol{0}$ instanton segment to a local approximation of the unstable manifold of $(\boldsymbol{x}_i^*,\boldsymbol{0})$, and the final point to a local approximation of the stable manifold of the intermediate saddle $(\boldsymbol{x}_\nu^*,\boldsymbol{0})$; details are given in  Appendix~\ref{app:locapp}. The final constraint fixes the initial displacement in state space, $\Vert\boldsymbol{r}_i(0) \Vert_2=\epsilon$, where $\boldsymbol{r}_i \equiv \boldsymbol{x}-\boldsymbol{x}^*_i$ and $\Vert\boldsymbol{a}\Vert_2\equiv\sqrt{\boldsymbol{a}\cdot\boldsymbol{a}}$ is the Euclidean norm. The solver then determines both the shape of instanton trajectory and its duration $T$. Unlike Ref.~\cite{lee2024real}, which fixes $T$, leaves the initial displacement $\epsilon$ free, and uses a linear local-manifold approximation, our implementation determines $T$, fixes $\epsilon$, and uses the refined local-manifold conditions described in Appendix~\ref{app:locapp}. The construction
of initial guesses for the collocation trajectory and its
duration $T$ is discussed in Appendix~\ref{app:initcol}.

\subsubsection{Shooting method}

Our second approach employs a shooting method. We integrate Hamilton's equations~\eqref{eq:hamilton-eq} from a point close to $(\boldsymbol{x}_i^*,\boldsymbol{0})$ on a local approximation to its unstable manifold. We then vary the initial condition and trajectory duration $T$ to minimize the terminal distance from the target saddle,
$\delta=\Vert\boldsymbol{q}_\nu(T)\Vert_2$, where $\boldsymbol{q}_\nu\equiv (\boldsymbol{x}-\boldsymbol{x}_\nu^*,\boldsymbol{\pi})$.
In particular, for both models studied here, the effective state-space dimension is $d=2$. For an escape from a spiral attractor (defined below), the initial-condition search therefore reduces to varying a single angular parameter specifying a direction in the two-dimensional unstable eigenspace of $(\boldsymbol{x}_i^*,\boldsymbol{0})$; see Appendix~\ref{app:locapp} for details.

\subsubsection{Methods comparison}
For the effectively two-dimensional ($d=2$) instanton problems studied here, collocation and shooting were most effective in complementary regimes. Collocation was preferable for trajectories escaping nodal attractors, whose mean-field Jacobian~\eqref{eq:mfjacobian} has two real, negative eigenvalues; shooting was less effective in this case, especially when their magnitudes were widely separated. By contrast, shooting was more effective for trajectories escaping spiral attractors, whose mean-field Jacobian~\eqref{eq:mfjacobian} has a complex-conjugate pair of eigenvalues with negative real parts. Appendix~\ref{app:initcol} discusses these differences in terms of the initial guess needed for collocation and the sensitivity of shooting to initial conditions.

\section{Strong symmetry: LMG model} \label{sec:lmg}
\subsection{Model}
Our first example illustrating the instanton framework is the driven-dissipative version of the Lipkin-Meshkov-Glick (LMG) model, which serves as a paradigmatic model of multistability in systems with strong permutation symmetry~\cite{morrison2008dynamicalprl,morrison2008collective,ferreira2019lipkin,wang2021dissipative,song2023crossover}. It has been microscopically derived as an effective model describing the collective dynamics of atoms in an optical cavity generated by laser-induced Raman transitions~\cite{morrison2008dynamicalprl,morrison2008collective}. An alternative implementation, based on magnetic atoms deposited on a metallic surface and driven by a spin-polarized current, has also been proposed~\cite{ferreira2019lipkin}. Specifically, we focus on the Ising-like version of this model, whose mean-field phase diagram and finite-size behavior have been thoroughly analyzed in Ref.~\cite{wang2021dissipative}. The model is described by the Lindblad master equation
\begin{align} \label{eq:masteqising}
\partial_t \hat{\rho}=\mathcal{L} \hat{\rho} \equiv &-i \left[\hat{H},\hat{\rho} \right] +\frac{\Gamma}{J} \mathcal{D} [\hat{J}_- ]\hat{\rho} \,,
\end{align}
with the Hamiltonian
\begin{align}
\hat{H}=\omega \hat{J}_z-\frac{\chi}{J} \hat{J}_x^2 \,.
\end{align}
Here, $\mathcal{L}$ is the Liouvillian superoperator, $\mathcal{D}[\hat{A}] \bullet\equiv\hat{A} \bullet \hat{A}^\dagger-\{\hat{A}^\dagger \hat{A},\bullet\}/2$ is the dissipator superoperator, $\hat{J}_\pm \equiv \hat{J}_x \pm i \hat{J}_y$ are ladder operators, $\omega$ is the effective magnetic field, $\chi$ is the interaction magnitude, and $\Gamma$ is the collective dissipation rate. We employ a Kac-like scaling of both the interaction and dissipation by $J$. This is merely a convenient convention that facilitates the analysis of the scaling behavior of the system with $J$; it does not imply that the interaction and dissipation physically weaken as $J$ increases. The numerical approach used to solve the above master equation is discussed in Appendix~\ref{app:masteq-lmg}.

\subsection{Mean-field dynamics}
In the limit of $J \rightarrow \infty$ the model is exactly described by nonlinear mean-field equations for the magnetization components $m_k \equiv\Tr[\hat{\rho} \hat{J}_k]/J$. They read as~\cite{wang2021dissipative} (see Appendix~\ref{app:meanfield})
 \begin{align} \nonumber
d_t m_x&=-\omega m_y +\Gamma m_x m_z \,, \\ \label{eq:meanfieldising}
d_t m_y &=\omega m_x+2 \chi m_x m_z +\Gamma m_y m_z \,, \\ \nonumber
d_t m_z &=-2 \chi m_x m_y-\Gamma (m_x^2+m_y^2) \,,
\end{align}
with $m_x^2+m_y^2+m_z^2=1$. To perform stability analysis of the fixed points, it is further convenient to rewrite the above equations in the stereographic coordinates $v \equiv m_x/(1-m_z)$, $w \equiv m_y/(1-m_z)$. This yields
\begin{subequations} \label{eq:lmgmeanfieldster}
\begin{align}
d_t v &=h_v =-\frac{4 \chi v^2 w}{v^2+w^2+1}-\Gamma  v-\omega w \,, \\
d_t w &=h_w =\chi \left(\frac{4 v^3}{v^2+w^2+1}-2 v \right)+\omega v -\Gamma  w \,.
\end{align}
\end{subequations} 

We further focus on the regime $\omega,\chi > 0$. In this regime, the system can relax to two types of stable fixed points (FPs) $m_k=m_k^*$. The first type of stable fixed points is the $\ell$ (lower) point with $m_z^*=-1$ and $m_x^*=m_y^*=0$. The second type consists of two $u_\pm$ (upper) FPs with $-1<m_z^*<0$. Due to the system $C_2$ symmetry (see Appendix~\ref{app:masteq-lmg}) these FPs are related by a rotation of 180$^\circ$ around the $z$-axis. They are associated with the magnetization components~\cite{wang2021dissipative}
\begin{subequations}
\begin{align}
m_x^* &=\pm \sqrt{(B_+-\omega^2 B_-/\Gamma^2)/(2\chi)} \,, \\
m_y^* &=\Gamma m_x^* m_z^*/\omega \,, \\
m_z^*&= -\omega/B_+ \,,
\end{align}
\end{subequations}
where
\begin{align}
B_{\pm} \equiv \chi \left(1 \pm \sqrt{1-\Gamma^2/\chi^2} \right) \,.
\end{align}
\begin{figure}
    \centering
    \includegraphics[width=0.96\linewidth]{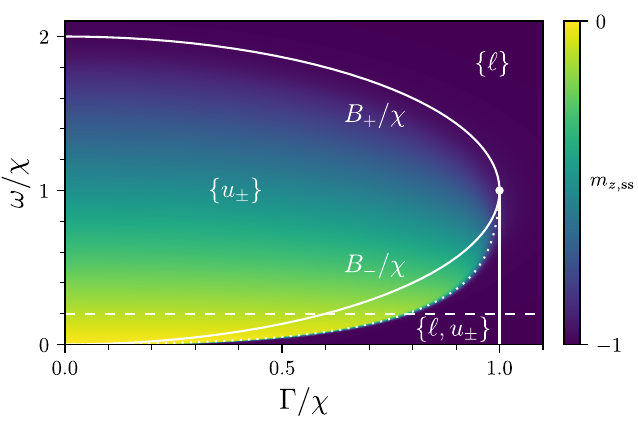}
    \caption{Phase diagram of the LMG model. Solid lines separate regions with different sets of stable fixed points, indicated by the symbols in curly braces. In the coexistence region of $\ell$ and $u_\pm$, the dotted line denotes the first-order phase-transition line determined by the crossing of the activation barriers, $\mathcal{A}_{\ell\to u_\pm}=\mathcal{A}_{u_\pm\to\ell}$ obtained using H representation. The dashed line marks the parameter-space path used in Fig.~\ref{fig:lmg-mzaction}. The color scale shows the steady-state magnetization $m_{z,\mathrm{ss}}$ obtained from the master equation for $J=200$.
    }
    \label{fig:lmg-bifurcation}
\end{figure}%
Figure~\ref{fig:lmg-bifurcation} presents the phase diagram, illustrating the parameter regions in which the above points are stable. As shown, the lower FP $\ell$ loses stability in the region $\Gamma/\chi<1$ and $B_-<\omega<B_+$. The upper FPs $u_\pm$ are stable in the region $\Gamma/\chi<1$ and $\omega < B_+$. Our analysis will focus on the region $\Gamma/\chi<1$ and $\omega < B_-$, where both types of FPs are stable.

\subsection{Instanton analysis}
Our goal now is to determine activation barriers $\mathcal{A}_{i \to j}$ for transitions between FPs $i,j \in \{\ell,u_\pm \}$. To that end, we represent the density matrix using the Husimi (H) and P representations $p_\text{H}$ and $p_\text{P}$ defined as~\cite{narducci1975multitime,altland2012quantum,mandt2015stochastic,burkle2020probabilistic}
\begin{subequations}
\begin{align}
&p_\text{H}(\theta,\varphi)  \equiv \tfrac{2J+1}{4\pi} \langle \theta,\varphi |\hat{\rho} |\theta,\varphi \rangle \,, \\
&\hat{\rho} \equiv \int  p_\text{P}(\theta,\varphi) |\theta,\varphi \rangle \langle \theta,\varphi |d\theta d\varphi \,,
\end{align}
\end{subequations}
where $|\theta,\varphi \rangle$ are coherent spin states~\cite{radcliffe1971some}
\begin{align}
&|\theta,\varphi \rangle \equiv \\\nonumber  &\sum_{M=-J}^J \sqrt{2J \choose J+M} \left( \cos\frac{\theta}{2} \right)^{J+M} \left( \sin\frac{\theta}{2} e^{i \varphi} \right)^{J-M} |J,M \rangle \,,
\end{align}
and $|J,M \rangle$ are eigenstates of $\hat{J}_z |J,M \rangle=M|J,M \rangle$. It is further convenient to move from spherical coordinates $\theta$, $\varphi$ to the stereographic coordinates $v$, $w$ using relation $\cot(\theta/2)e^{i \varphi}=v+iw$, so that the vector $\boldsymbol{x}$ in Eq.~\eqref{eq:evquas} is defined as $\boldsymbol{x} \equiv (v,w)$. The derivation of differential operators $L_\alpha$ describing the evolution of $p_\alpha$ for $\alpha \in \{\text{H},\text{P}\}$ is discussed in Appendix~\ref{app:auxhamlmg-derhp}, and the explicit form of the resulting auxiliary Hamiltonians $\mathcal{H}_\alpha$ is presented in Appendix~\ref{app:auxhamlmg-res}.

\begin{figure}
    \centering
    \includegraphics[width=0.96\linewidth]{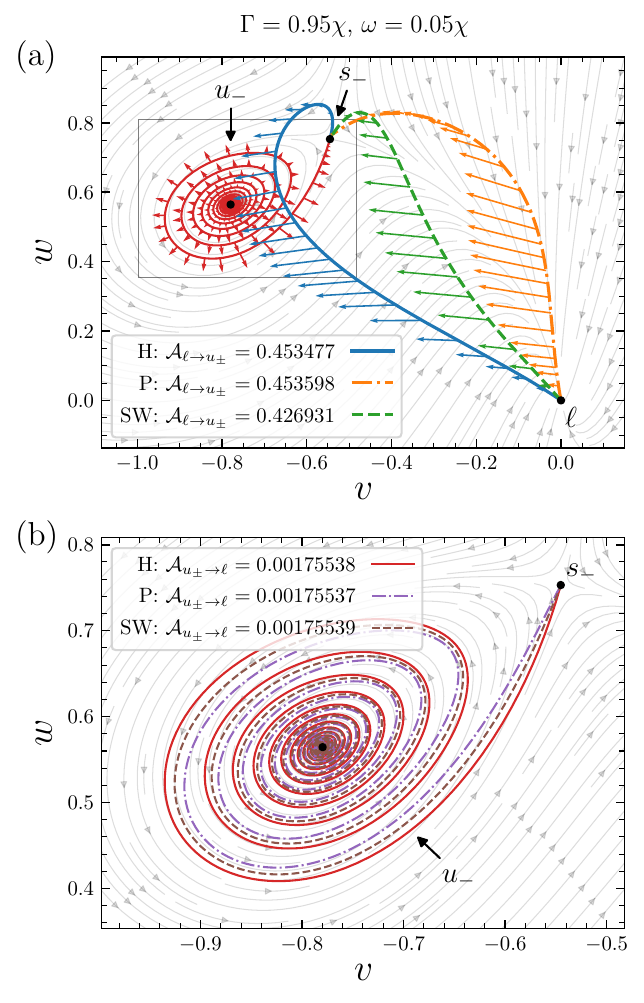}
    \caption{Projections of the $\boldsymbol{\pi}\neq\boldsymbol{0}$ segments of instanton trajectories onto the $(v,w)$ state plane for $\Gamma=0.95\chi$ and $\omega=0.05\chi$. (a) Three instanton trajectories connecting the fixed point $\ell$ to the saddle $s_-$, obtained for different quasiprobability representations, together with a single trajectory connecting the fixed point $u_-$ to $s_-$ for the H representation. The arrows attached to the trajectories indicate the corresponding momentum vectors $\boldsymbol{\pi}$. (b) Instanton trajectories connecting $u_-$ to $s_-$ for different representations, shown for the region indicated by the gray rectangle in panel (a). Momentum arrows are omitted in (b) for clarity. In both panels, the gray background lines with arrows represent the mean-field velocity field.
    }
    \label{fig:lmg-instantons}
\end{figure}%

The instanton trajectories connecting stable FPs $i,j \in \{\ell, u_\pm \}$ are composed of two segments. The first segment, associated with $\boldsymbol{\pi} \neq \boldsymbol{0}$ and positive action, connects FP $i$ to a saddle point $s_{\pm}$ with magnetization components~\cite{wang2021dissipative}
\begin{subequations}
\begin{align}
m_x^* &=\pm \sqrt{(B_--\omega^2 B_+/\Gamma^2)/(2\chi)} \,, \\
m_y^* &=\Gamma m_x^* m_z^*/\omega \,, \\
m_z^*&= -\omega/B_- \,.
\end{align}
\end{subequations}
Then, the system follows a mean-field relaxation to the FP $j$, which is associated with $\boldsymbol{\pi}=\boldsymbol{0}$, and thus zero action. The representative $\boldsymbol{\pi} \neq \boldsymbol{0}$ segments of the instanton trajectories are presented in Fig.~\ref{fig:lmg-instantons}. The H and P representations yield markedly different instanton trajectories, yet their actions agree within numerical accuracy. The small residual differences reflect the approximation of infinite-time instantons by finite-time numerical trajectories (see Sec.~\ref{subsec:framework-num}). As shown in Fig.~\ref{fig:lmg-instantons}, trajectories escaping the nodal fixed point $\ell$ follow smooth arcs, whereas those escaping the spiral fixed point $u_-$ spiral away from it. The numerical algorithm converges for nearly all parameter values studied in the H representation, but fails to converge for some in the P representation.

\subsection{Magnetization behavior}
\begin{figure}
    \centering
    \includegraphics[width=0.95\linewidth]{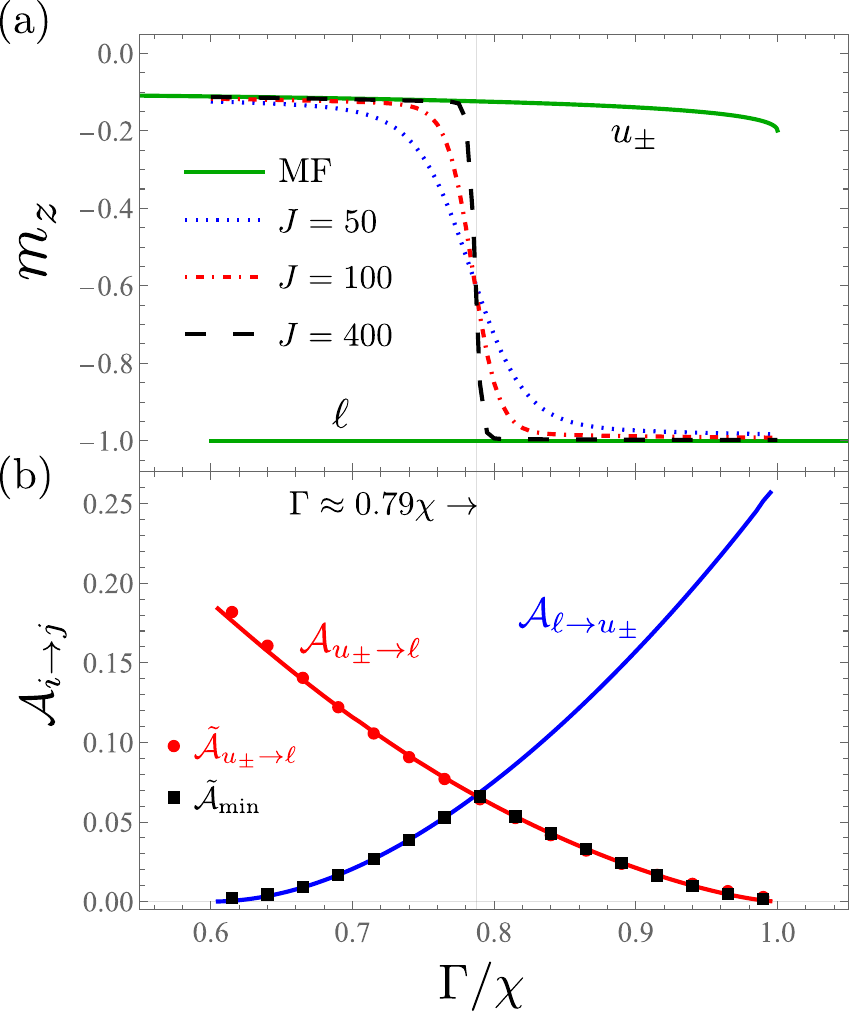}
    \caption{(a) Mean-field (MF) fixed-point magnetization $m_z^*$ (green solid lines) versus finite-size steady-state magnetization $m_{z,\text{ss}}$ obtained using master equation (other lines). (b) Activation barriers $\mathcal{A}_{i \rightarrow j}$ calculated using H representation; they agree with the results obtained using P representation whenever the latter method converges. Grey vertical solid line at $\Gamma \approx 0.79 \chi$ denotes the crossing point of $\mathcal{A}_{\ell \rightarrow u_\pm}$ and $\mathcal{A}_{u_\pm \rightarrow \ell}$ curves. Symbols denote the estimators $\tilde{\mathcal{A}}_{u_\pm \to \ell}$ and $\tilde{\mathcal{A}}_{\text{min}}$ defined by Eq.~\eqref{eq:lmg-actionest}. Parameter $\omega=0.2 \chi$.
    }
    \label{fig:lmg-mzaction}
\end{figure}

To demonstrate the validity of the instanton approach, we analyze the steady-state magnetization at large but finite $J$ in the parameter region where the mean-field dynamics predicts the coexistence of the $\ell$ and $u_\pm$ attractors. In contrast to the mean-field description, a finite system admits a unique stationary state $\mathcal{L} \hat{\rho}_\text{ss}=0$. The corresponding magnetization $m_{z,\text{ss}} \equiv \Tr[\hat{\rho}_\text{ss} \hat{J}_z]/J$ is plotted in Fig.~\ref{fig:lmg-mzaction}(a) and compared with the magnetization components $m_z^*$ of the mean-field attractors. For $\Gamma \lessapprox 0.79 \chi$, the stationary magnetization remains close to the $u_\pm$ branch of the mean-field solution. Near $\Gamma \approx 0.79 \chi$, however, it rapidly departs from this branch and approaches the $\ell$ branch, with the crossover becoming progressively sharper as $J$ increases.

This behavior can be understood from the finite-size switching dynamics between the metastable states associated with the mean-field attractors. At finite $J$, these states have long but finite lifetimes, and their stationary occupations are determined by the corresponding switching rates. Since $J$ is the extensive parameter $V$ here, Eq.~\eqref{eq:arrh} gives the ratio of probabilities of the $u_\pm$ and $\ell$ branches as $P_{u_\pm}/P_\ell=\kappa_{\ell \rightarrow u_\pm}/\kappa_{u_\pm \rightarrow\ell} \asymp \exp[J (\mathcal{A}_{u_\pm \rightarrow \ell}-\mathcal{A}_{\ell \rightarrow u_\pm})]$. For $\Gamma \lessapprox 0.79 \chi$, $\mathcal{A}_{u_\pm \rightarrow \ell}>\mathcal{A}_{\ell \rightarrow u_\pm}$, so escape from the $u_\pm$ attractors is exponentially more strongly suppressed than escape from the $\ell$ attractor as $J$ increases. Consequently, the joint occupation of the $u_\pm$ states tends to one. This picture is confirmed by Fig.~\ref{fig:lmg-mzaction}(b), which shows the activation barriers $\mathcal{A}_{i \to j}$ obtained using the instanton approach. For $\Gamma \gtrapprox 0.79 \chi$, the ordering of the activation barriers is reversed, and the $\ell$ state becomes dominant. In the limit $J \rightarrow \infty$ (taken after the $t \rightarrow \infty$ limit), the finite-size crossover therefore sharpens into a first-order dissipative phase transition at the point where $\mathcal{A}_{u_\pm \rightarrow \ell}=\mathcal{A}_{\ell \rightarrow u_\pm}$, with $m_{z,\text{ss}}$ jumping discontinuously from the $u_\pm$ to the $\ell$ branch.

The instanton approach therefore provides a direct prediction for the location of the first-order dissipative phase transition from the crossing of the activation barriers. This is further illustrated in Fig.~\ref{fig:lmg-bifurcation}, which shows the stationary magnetization as a function of $\Gamma$ and $\omega$. The white dotted line in the coexistence regions of $\ell$ and $u_\pm$ attractors denotes the parameter values at which the activation barriers become equal, $\mathcal{A}_{u_\pm \rightarrow \ell}=\mathcal{A}_{\ell \rightarrow u_\pm}$. The close agreement between this line and the sharp variation of the stationary magnetization shows that the instanton analysis accurately identifies the first-order transition boundary throughout the parameter space.

\subsection{Liouvillian gap scaling}\label{sec:liovgap-lmg}
Beyond characterizing the steady state, the instanton approach also allows us to determine the exponential scaling of the relaxation rates associated with switching between metastable states. The relevant quantity is the Liouvillian gap, which sets the slowest relaxation timescale of the system. To define it, we order the eigenvalues $\lambda_i$ of $\mathcal{L} \hat{\rho}_i=\lambda_i \hat{\rho}_i$ according to $0=- \re \lambda_0 \leq -\re \lambda_1\leq \ldots$. The Liouvillian gap is then defined as $\lambda \equiv -\re \lambda_1$~\cite{minganti2018spectral}.

\begin{figure}
    \centering
    \includegraphics[width=0.95\linewidth]{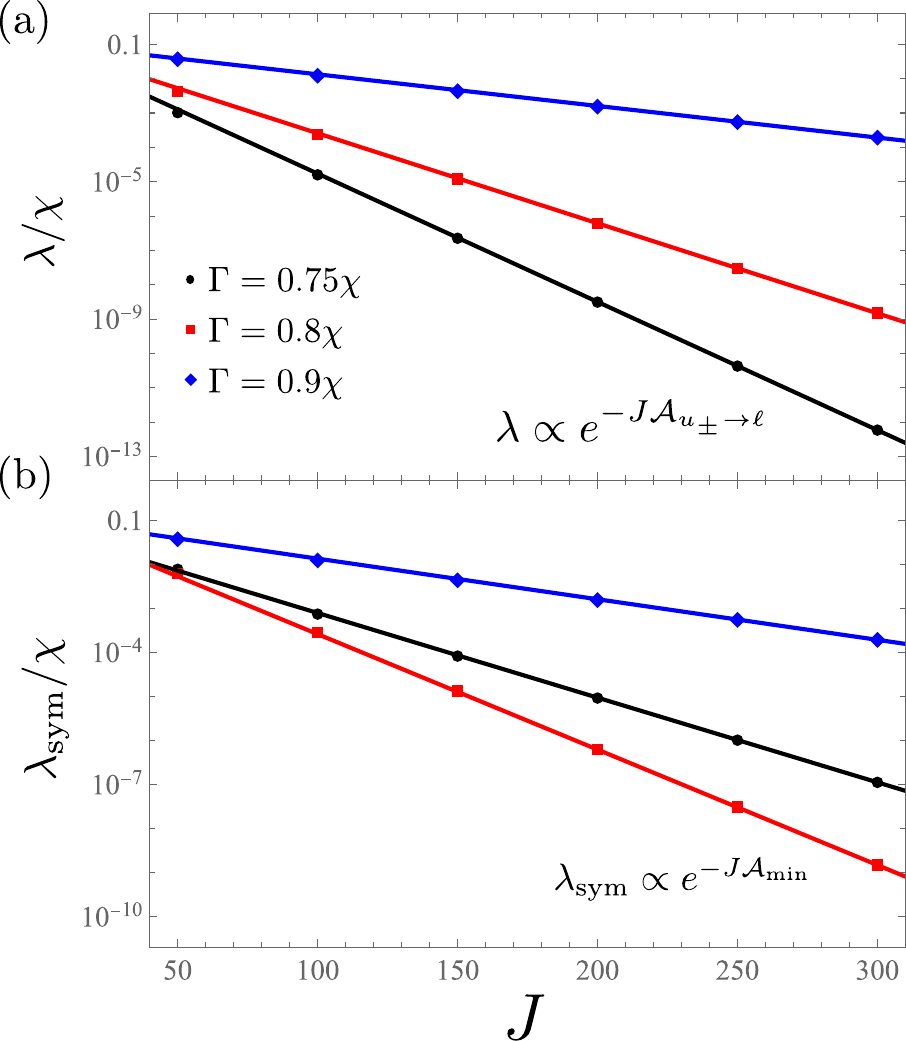}
    \caption{(a) Dots: Liouvillian gap $\lambda$ calculated using master equation for selected values of $\Gamma$, plotted in the logarithmic scale. Lines: the predicted scaling $\lambda \propto \exp(-J \mathcal{A}_{u_\pm \to \ell})$ set to cross the point for $J=300$. (b) An analogous plot for the symmetric Liouvillian gap $\lambda_\text{sym}$, with lines representing the scaling $\lambda_\text{sym} \propto \exp(-J \mathcal{A}_\text{min})$.
    }
    \label{fig:lmg-eigenvalues}
\end{figure}

In the case considered here, the Liouvillian gap is always controlled by the escape rate from the $u_\pm$ attractor, $\lambda \approx \kappa_{u_\pm \rightarrow \ell}$, although the physical character of the corresponding slow mode changes across the point where the relative stability of the metastable states is reversed. When the $u_\pm$ attractors are less probable, the slowest mode describes the decay of their total population toward the dominant $\ell$ state. When the $u_\pm$ attractors are more probable, it instead describes the relaxation of the population imbalance between the $u_+$ and $u_-$ attractors. This imbalance breaks the system $C_2$ symmetry and relaxes through switching between $u_+$ and $u_-$ via the intermediate $\ell$ attractor. In both cases, the slowest process is therefore limited by escape from $u_\pm$, $\lambda \approx \kappa_{u_\pm \rightarrow \ell}$, which by Eq.~\eqref{eq:arrh} yields the asymptotic scaling $\lambda \asymp e^{-J\mathcal{A}_{u_\pm \to \ell}}$. This exponential dependence is confirmed in Fig.~\ref{fig:lmg-eigenvalues}(a) for several representative parameter values.

The $C_2$ symmetry also allows us to separate symmetric and antisymmetric relaxation modes. The eigenstates $\hat{\rho}_i$ can be classified according to their behavior under a rotation by 180$^\circ$ around the $z$-axis, with the population-imbalance mode between $u_+$ and $u_-$ belonging to the odd sector (see Appendix~\ref{app:masteq-lmg}). This motivates the definition of the symmetric Liouvillian gap $\lambda_\text{sym}$, obtained by restricting the spectrum to eigenstates that are even under such a rotation. Since this excludes the $u_+$--$u_-$ imbalance mode, $\lambda_\text{sym}$ is controlled by relaxation from the less probable to the more probable attractor, $\lambda_\text{sym} \approx \max\{\kappa_{\ell \to u_\pm},\kappa_{u_\pm \to \ell} \}$. Consequently, it obeys the asymptotic scaling $\lambda_\text{sym} \asymp e^{-J \mathcal{A}_\text{min}}$, where $\mathcal{A}_\text{min} \equiv \min\{\mathcal{A}_{\ell \rightarrow u_\pm},\mathcal{A}_{u_\pm \rightarrow \ell}\}$, as demonstrated in Fig.~\ref{fig:lmg-eigenvalues}(b).

Using these exponential scalings, we can also construct estimators of the corresponding activation barriers directly from the Liouvillian and symmetric Liouvillian gaps,
\begin{subequations} \label{eq:lmg-actionest}
\begin{align}
\tilde{\mathcal{A}}_{u_\pm \to \ell} &\equiv \frac{1}{10} \ln \frac{\lambda(J=120)}{\lambda(J=130)} \approx \mathcal{A}_{u_\pm \to \ell} \,, \\ \label{eq:lmg-amin}
\tilde{\mathcal{A}}_\text{min} &\equiv \frac{1}{100} \ln \frac{\lambda_\text{sym}(J=300)}{\lambda_\text{sym}(J=400)} \approx \mathcal{A}_\text{min} \,.
\end{align}
\end{subequations}
These estimators are shown as symbols in Fig.~\ref{fig:lmg-mzaction}(b). Their close agreement with the activation barriers obtained from the instanton approach provides an independent confirmation of the predicted exponential scaling and of the validity of the instanton description.

\subsection{Manifestations of Hamiltonian nonconvexity} \label{subsec:lmgnonconvex}
Having established its physical relevance, we now turn to some technical and formal aspects of the instanton framework. For the model considered here, the quasiprobability representations $\alpha \in \{\text{H},\text{P}\}$ yield operators $L_{\alpha}$ of the Fokker--Planck form (Appendix~\ref{app:auxhamlmg-res}). However, these operators do not define genuine classical stochastic dynamics, since the corresponding diffusion matrices are not positive semidefinite. Consequently, the associated auxiliary Hamiltonians $\mathcal{H}_{\alpha}$ are nonconvex in the momenta $\pi_{v,w}$. This is in contrast to the instanton framework developed for metastability in classical stochastic systems, where the auxiliary Hamiltonians are convex. In particular, the lack of convexity precludes the direct application of efficient minimum-action algorithms developed for classical systems with convex Hamiltonians~\cite{grafke2019numerical,zakine2023minimum,gagrani2023action}. 

We now discuss how this nonconvexity manifests itself in the associated instanton trajectories and, more specifically, in the behavior of the action
\begin{align} \nonumber
&S_\alpha(\boldsymbol{x}'',t|\boldsymbol{x'},0) = \\ \label{eq:action} &\int_0^t dt' \left\{ \boldsymbol{\pi}(t') d_{t'} \boldsymbol{x}(t')-\mathcal{H}_\alpha[\boldsymbol{x}(t'),\boldsymbol{\pi}(t')] \right\} \,.
\end{align}
For both convex and nonconvex Hamiltonians the activation barriers are determined as $\mathcal{A}_{i \to j}=\lim_{T \to \infty} S_\alpha(\boldsymbol{x}_j^*,T|\boldsymbol{x}_i^*,0)$, calculated for evolution along the instanton trajectory obeying the boundary conditions $\boldsymbol{\pi}(0)=\boldsymbol{\pi}(T)=\boldsymbol{0}$. However, for convex Hamiltonians the action additionally obeys the following properties~\cite{grafke2019numerical,zakine2023minimum,gagrani2023action}:
\begin{enumerate}[label=(\alph*)]
 \item The action $S_\alpha(\boldsymbol{x}'',t|\boldsymbol{x}',0)$ is always nonnegative.
 \item The action $S_\alpha(\boldsymbol{x}'',t|\boldsymbol{x}',0)$ is nondecreasing ($d_t S_\alpha \geq 0$).
\item The activation barrier $\mathcal{A}_{i \to j}$ corresponds to a minimum action $S_\alpha(\boldsymbol{x}_j^*,t|\boldsymbol{x}_i^*,0)$ minimized over $t$, $\boldsymbol{\pi}(0)$, and $\boldsymbol{\pi}(t)$.
\end{enumerate}
\begin{figure}
    \centering
    \includegraphics[width=0.96\linewidth]{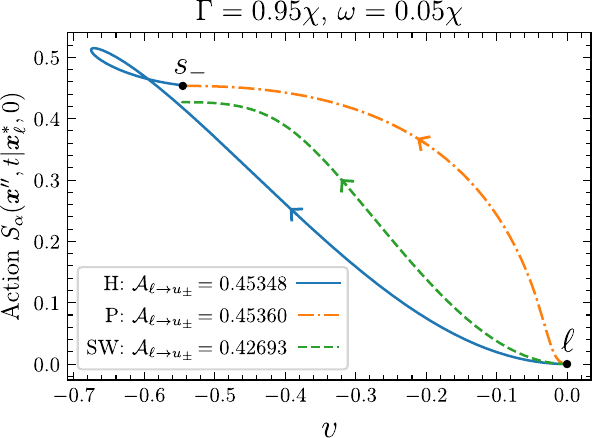}
    \caption{Transient evolution of the action $S_\alpha(\boldsymbol{x}'',t|\boldsymbol{x}_\ell^*,0)$ along the instanton segment connecting the fixed point $\ell$ to the saddle $s_-$ shown in Fig.~\ref{fig:lmg-instantons}(a), for different quasiprobability representations. The action is plotted parametrically versus the $v$ coordinate.
    Parameters are as in Fig.~\ref{fig:lmg-instantons}: $\Gamma=0.95\chi$ and $\omega=0.05\chi$.
    }
    \label{fig:lmg-transient-action}
\end{figure}%
As observed in the companion Letter~\cite{ptaszynski2026quantum}, all these conditions can be broken for systems with nonconvex Hamiltonian. For the LMG model, we did not observe breaking of the condition (a), though we also cannot rule it out. The possibility of breaking of the condition (c) is more difficult to analyze due to more complex instanton geometry than in the model from the companion Letter. However, we observed breaking of the condition (b). This is illustrated in Fig.~\ref{fig:lmg-transient-action} which presents the transient evolution of the action $S_\alpha$ along the instanton segment connecting points $\ell$ and $s_-$.
As shown, the actions $S_\alpha$ converge to the same final value for $\alpha \in \{\text{H},\text{P}\}$, but have a different transient behavior. In particular, for the H representation, the action decreases near the end of trajectory, breaking condition (b).

\subsection{Comparison with semiclassical Wigner approach} \label{subsec:lmg-sw}
The convex auxiliary Hamiltonians are instead provided by a different method, referred to as the semiclassical Wigner (SW) approach. The SW approach starts from the differential operator governing the evolution of the Wigner distribution and truncates it at second order in derivatives~\cite{dubois2021semi,gronchi1979fokker}. The resulting Fokker--Planck generator is positive definite and therefore admits a mapping onto a classical stochastic process. However, as demonstrated for the model discussed in the companion Letter~\cite{ptaszynski2026quantum}, this procedure generally does not ensure an accurate characterization of activation barriers $\mathcal{A}_{i \to j}$. The reason is that it relies on an uncontrolled truncation of the higher-derivative terms, which---according to Eq.~\eqref{eq:auxhamdef}---contribute to the auxiliary Hamiltonian $\mathcal{H}_{\alpha}$ at the same order as the first- and second-derivative terms.

The situation is markedly different for the LMG model analyzed in the present work (see implementation details of SW approach in Appendix~\ref{app:auxhamlmg-dersw} and its final form in Appendix~\ref{app:auxhamlmg-res}). Although the activation barriers predicted by the SW approach differ from those obtained using the representations $\alpha \in \{\text{H},\text{P}\}$, these discrepancies are nearly negligible in most of the parameter range considered.  Indeed, if the activation barriers predicted by the SW approach were included in Fig.~\ref{fig:lmg-mzaction}(b), they would be visually indistinguishable from those obtained using the H and P representations. In particular, we do not identify any regime in which the activation barrier values predicted by the SW approach lead to an appreciably incorrect description of observable quantities, such as stationary magnetization $m_{z,\text{ss}}$ or scaling of the Liouvillian gap.

Noticeable discrepancies arise only in the small range of parameter values, for the activation barriers $\mathcal{A}_{\ell \to u_{\pm}}$ in the strongly asymmetric regime $\mathcal{A}_{\ell \to u_{\pm}} \gg \mathcal{A}_{u_{\pm} \to \ell}$, as presented in Figs.~\ref{fig:lmg-instantons} and~\ref{fig:lmg-transient-action}. In this regime, the relative difference between the values of $\mathcal{A}_{\ell \to u_{\pm}}$ predicted by the SW approach and using $\alpha \in \{\text{H},\text{P}\}$ representations can become substantial. Nevertheless, this discrepancy does not produce an appreciable difference in the behavior of the observable quantities considered here, since these quantities are controlled by the smaller activation barrier $\mathcal{A}_{u_{\pm} \to \ell}$. Thus, for the parameter regime studied here, the SW approach yields quantitatively accurate predictions for the metastable behavior of the system despite providing inaccurate values of some individual activation barriers. This agreement should, however, be regarded as model- and parameter-dependent rather than as evidence for the general validity of the approximation. The SW approach relies on an uncontrolled truncation that is not formally justified and---as explicitly demonstrated for the model considered in the companion Letter~\cite{ptaszynski2026quantum} and the CRF model in Sec.~\ref{sec:crf}---can yield quantitatively incorrect activation barriers $\mathcal{A}_{i \to j}$ that result in an incorrect description of the observable quantities governed by them.

\section{Weak symmetry: CRF model} \label{sec:crf}
\subsection{Model}
Our second example is the cooperative resonance fluorescence (CRF) model, which has been used to model an experimental setup consisting of an ensemble of $N$ Sr atoms confined in an optical cavity in the bad-cavity regime~\cite{song2025dissipation}. It is described by the Lindblad equation (in the rotating wave approximation)
\begin{align} \nonumber
\partial_t \hat{\rho}=&-i \left[\Omega \hat{J}_x,\hat{\rho} \right]  +\frac{\Gamma}{N} \mathcal{D} [\hat{J}_- ]\hat{\rho} \\&+ \sum_{i=1}^N \left( \gamma \mathcal{D} [\sigma_-^i] \hat{\rho}+\frac{\kappa}{2} \mathcal{D}[\sigma_z^i]\hat{\rho} \right) \,.
\end{align}
Here, $\sigma^i_\pm=(\sigma^i_x \pm i \sigma^i_y)/2$ are local ladder operators, $\Omega$ is the amplitude of the drive resonant with the cavity frequency, $\Gamma$ is the collective dissipation rate (corresponding to dissipation mediated by an optical cavity), $\gamma$ is the local dissipation rate, and $\kappa$ is the local dephasing rate. As in the LMG model from the previous section, we apply the Kac-like scaling of the collective dissipation term by the number of atoms $N$. Compared to the LMG model, the last two terms of the Lindblad equation act identically but independently on each atom, preserving weak permutation symmetry. As a consequence, the total angular momentum is not conserved. A numerical framework used to solve this master equation is presented in Appendix~\ref{app:master-crf}.

\subsubsection{Mean-field dynamics and instanton analysis}
Analogously to the LMG model, in the limit of $N \rightarrow \infty$ the model is exactly described by nonlinear mean-field equations for the magnetization components, now defined as $m_k \equiv 2\Tr[\hat{\rho} \hat{J}_k]/N$. They read as~\cite{leppenen2026quantum} (see Appendix~\ref{app:meanfield})
 \begin{align} \nonumber
d_t m_x&=\Gamma m_x m_z/2 -\gamma m_x/2-\kappa m_x \,, \\ \label{eq:meanfieldcrf}
d_t m_y &=-\Omega m_z+\Gamma m_y m_z/2-\gamma m_y/2-\kappa m_y \,, \\ \nonumber
d_t m_z &=\Omega m_y-\Gamma (m_x^2+m_y^2)/2-\gamma(1+m_z) \,.
\end{align}
In certain parameter regimes, this dynamics has two stable fixed points, which we discuss in detail later. For the instanton analysis, we use two kinds of quasiprobability representations. The first one is the normal-order quasiprobability representation~\cite{haken1967quantum,carmichael1999statistical1}
\begin{align} \label{eq:nodistribution-normalized}
p_\text{NO} (\boldsymbol{m}) \equiv\frac{N^3}{2 \pi^3} \int d^2 \xi d \eta \Tr[\hat{\rho} \hat{O}_\text{NO} ] e^{-iN (\xi^* m_++\eta m_z+\xi m_-)} \,,
\end{align}
where $\boldsymbol{m}\equiv (m_x,m_y,m_z)$ is the magnetization vector, $m_\pm \equiv (m_x \pm im_y)/2$, and 
\begin{align}
\hat{O}_\text{NO} \equiv e^{i \xi^* \hat{J}_+} e^{2i \eta \hat{J}_z} e^{i \xi \hat{J}_-} \,.
\end{align}
The second one is the generalized coherent state representation which is implicitly defined as~\cite{merkel2021phase}
\begin{align}
\hat{\rho}=\int p_\text{GC}(\boldsymbol{m}) \hat{g}(\boldsymbol{m}) d\boldsymbol{m} \,,
\end{align}
where
\begin{align}
\hat{g}(\boldsymbol{m}) = \bigotimes_{i=1}^N \frac{1}{2} \left(\mathds{1}+\sum_{k \in \{x,y,z \}} m_k \sigma_k^i \right)
\end{align}
are generalized coherent states. The derivations of the corresponding auxiliary Hamiltonians are discussed in Appendix~\ref{app:auxhamcrf}, with the final result summarized in Appendix~\ref{app:auxhamcrf-res}. We note that the auxiliary Hamiltonians involve six variables, $m_{x,y,z}$ and $\pi_{x,y,z}$. However, we found that---due to system mirror symmetry with respect to $(m_y,m_z)$ plane---the fixed points of mean-field equations are associated with $m_x^*=0$, and the instanton trajectories connecting them are associated with $m_x=\pi_x=0$. This reduces the instanton problem to four dimensions, as for the LMG model.

\subsection{Zero dephasing ($\kappa=0$) case}
We now discuss the system behavior, starting from the zero dephasing case of $\kappa=0$. This regime admits an exact solution, corresponding to the zero-interaction case of the dissipative transverse-field Ising model explored in Ref.~\cite{roberts2023exact}. This exact solution provides a rigorous benchmark for the instanton approach.

\begin{figure}
    \centering
    \includegraphics[width=0.95\linewidth]{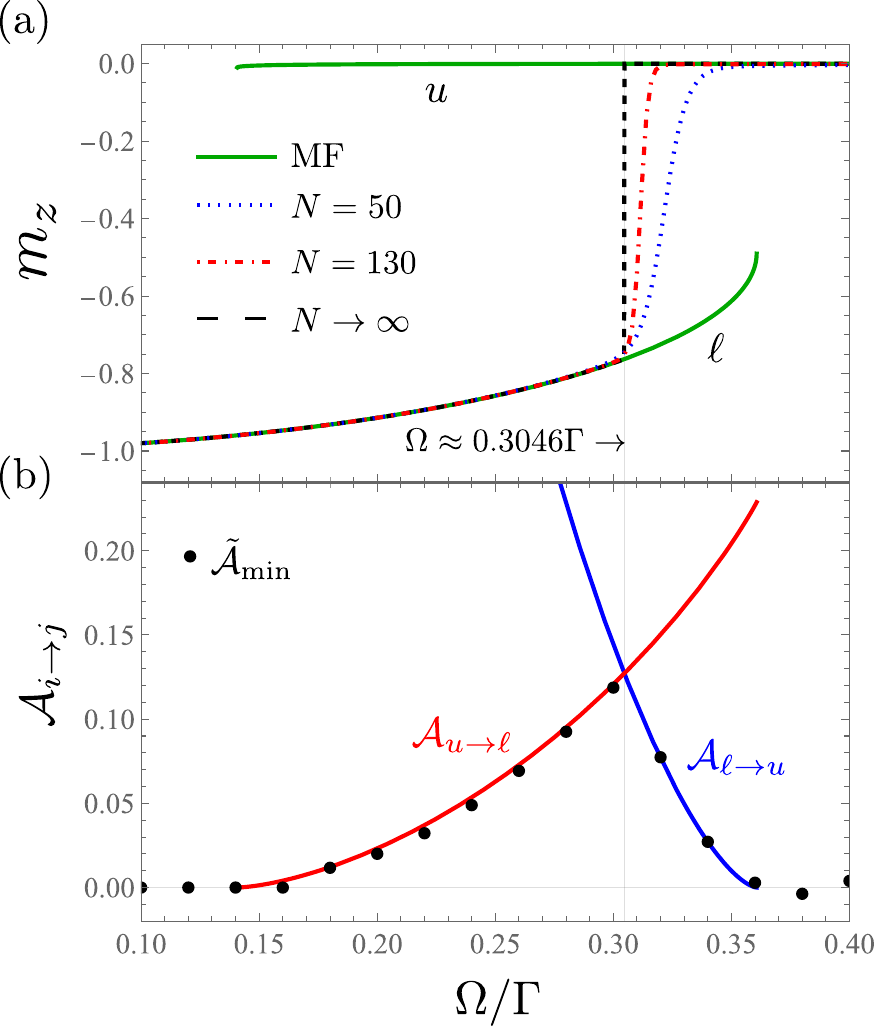}
    \caption{Results for the zero dephasing case $\kappa=0$. (a) Mean-field (MF) fixed-point magnetization $m_z^*$ (green solid lines) versus finite-size steady-state magnetization $m_{z,\text{ss}}$ obtained using master equation and exact steady-state magnetization for $N \to \infty$ (other lines). (b) Activation barriers $\mathcal{A}_{i \rightarrow j}$ calculated from the analytic expression~\eqref{eq:crf-barriers-analytic}; they agree with the results obtained using NO representation, as well as GC representation whenever the latter method converges. Grey vertical solid line at $\Omega \approx 0.3046 \Gamma$ denotes the crossing point of $\mathcal{A}_{\ell \rightarrow u}$ and $\mathcal{A}_{u \rightarrow \ell}$ curves. Symbols denote the estimator $\tilde{\mathcal{A}}_\text{min}$ defined by Eq.~\eqref{eq:crf-amin}. Parameters: $\gamma=0.01\Gamma$.
    }
    \label{fig:crf-mzaction-kappa0}
\end{figure}

As discussed in Ref.~\cite{leppenen2026quantum}, for $\gamma<\Gamma/8$ and within a certain range of $\Omega$, the mean-field dynamics exhibits two stable fixed points: the lower FP $\ell$, with $m_z^*\approx -1$, and the upper FP $u$, with $m_z^*\approx 0$. The corresponding fixed-point magnetizations $m_z^*$ are shown in Fig.~\ref{fig:crf-mzaction-kappa0}(a), where they are compared with the finite-size steady-state magnetization obtained from the master equation and with the exact result for the steady-state magnetization in the limit $N\to\infty$, taken after the long-time limit $t\to\infty$. According to the latter result, the stationary magnetization is given by~\cite{roberts2023exact}
\begin{align}
\lim_{N \rightarrow \infty} m_{z,\text{ss}}
=
\argmin_{m_z\in[-1,0]} V(m_z) \,,
\end{align}
with the effective potential
\begin{align}
V(m_z)
=
\ln \frac{|m_z|^{-m_z}(1+m_z)^{1+m_z}}
{(\gamma/\Gamma-m_z)^{2(\gamma/\Gamma-m_z)}}
+
m_z\left(\ln\frac{\Gamma^2}{2\Omega^2}-2\right) \,.
\end{align}

As shown in Fig.~\ref{fig:crf-mzaction-kappa0}(a), for the parameters considered, the system undergoes a first-order transition at $\Omega\approx0.3046\Gamma$, at which the steady-state magnetization $m_{z,\mathrm{ss}}$ jumps from the $\ell$ branch to the $u$ branch. For finite system sizes, the steady-state magnetization remains concentrated near the $\ell$ branch approximately up to this point and then crosses over toward the $u$ branch, with the crossover becoming sharper as $N$ increases.

Following the discussion of the LMG model, the position of the first-order transition can alternatively be determined from the crossing of the activation barriers,
$\mathcal{A}_{\ell\to u}=\mathcal{A}_{u\to\ell}$.
Moreover, as conjectured in Ref.~\cite{noachtar2026exact}, the activation barriers themselves can be obtained from the effective potential $V(m_z)$,
\begin{subequations}
\label{eq:crf-barriers-analytic}
\begin{align}
\mathcal{A}_{\ell \to u}
&=
V(m_{z,s}^*)-V(m_{z,\ell}^*) \,,\\
\mathcal{A}_{u \to \ell}
&=
V(m_{z,s}^*)-V(m_{z,u}^*) \,,
\end{align}
\end{subequations}
where $s$ denotes a saddle point of the mean-field equations. The resulting barriers are shown in Fig.~\ref{fig:crf-mzaction-kappa0}(b).

\begin{figure}
    \centering
    \includegraphics[width=0.95\linewidth]{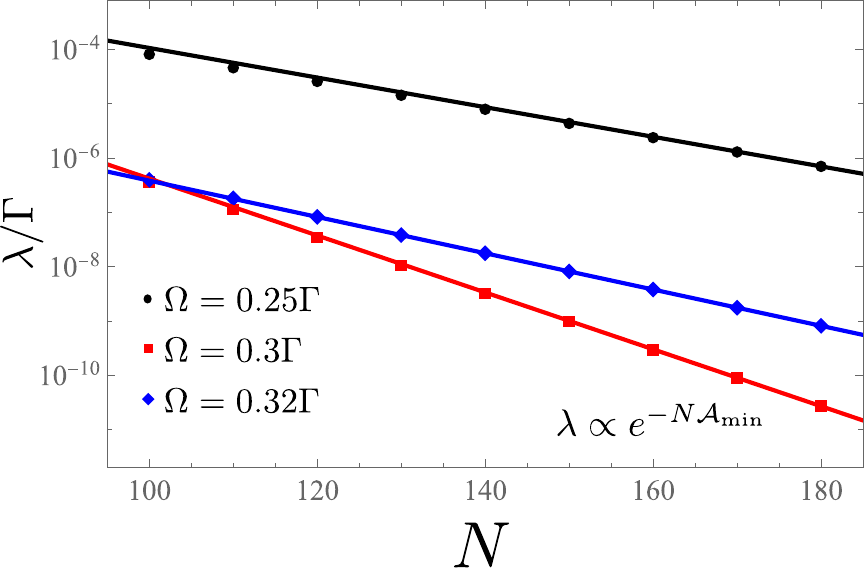}
    \caption{
    Results for the zero-dephasing case, $\kappa=0$.
    Dots show the Liouvillian gap $\lambda$ obtained from the master equation for selected values of $\Omega$ on a logarithmic scale.
    Lines indicate the predicted scaling
    $\lambda\propto\exp(-N\mathcal{A}_{\text{min}})$,
    normalized to pass through the corresponding data point at $N=180$.
    }
    \label{fig:crf-eigenvalues-kappa0}
\end{figure}%

In Ref.~\cite{noachtar2026exact}, this relation was verified numerically by comparison with the scaling of the Liouvillian gap. Here, we provide an independent confirmation by presenting the Liouvillian-gap scaling for representative parameter values in Fig.~\ref{fig:crf-eigenvalues-kappa0}. The numerical results agree well with the expected asymptotic behavior
$\lambda\asymp e^{-N\mathcal{A}_\text{min}}$,
where
$\mathcal{A}_\text{min}\equiv
\min\{\mathcal{A}_{\ell\to u},\mathcal{A}_{u\to\ell}\}$.
We additionally consider the estimator
\begin{align}
\label{eq:crf-amin}
\tilde{\mathcal{A}}_\text{min}
&\equiv
\frac{1}{10}
\ln\frac{\lambda(N=170)}{\lambda(N=180)} \approx \mathcal{A}_\text{min} \,,
\end{align}
defined analogously to Eq.~\eqref{eq:lmg-amin}. As shown in Fig.~\ref{fig:crf-mzaction-kappa0}(b), this estimator is in good agreement with $\mathcal{A}_\text{min}$ obtained from Eq.~\eqref{eq:crf-barriers-analytic}.

\begin{figure}
    \centering
    \includegraphics[width=0.96\linewidth]{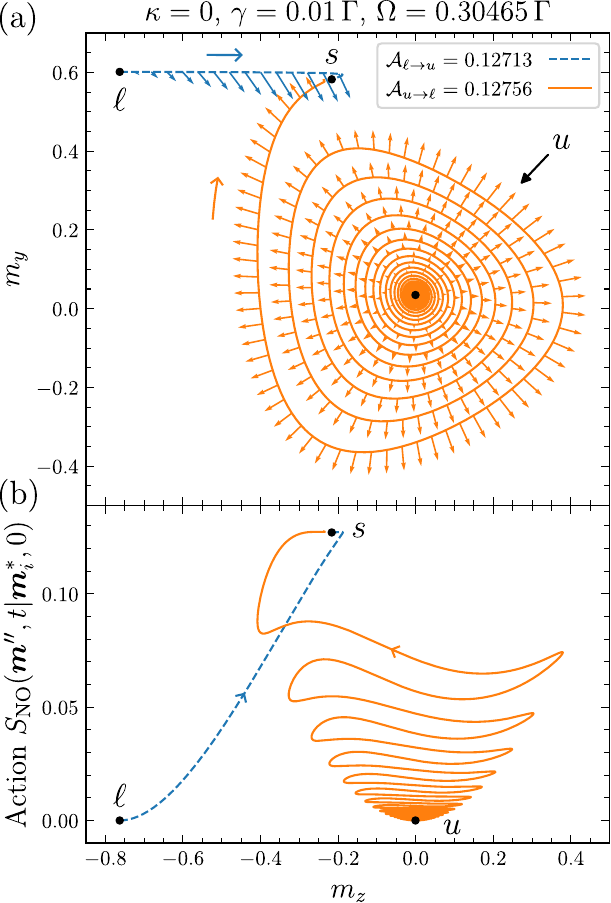}
    \caption{Results for the zero dephasing case $\kappa=0$ in the NO representation at the first-order phase-transition point, $\gamma=0.01\Gamma$ and $\Omega\approx0.3046\Gamma$.
(a) Projections onto the $(m_y,m_z)$ state plane of the $\boldsymbol{\pi}\neq\boldsymbol{0}$ instanton segments connecting the fixed points $\ell$ and $u$ to the saddle $s$. The arrows attached to the trajectories indicate the corresponding momentum vectors $\boldsymbol{\pi}$.
(b) Transient evolution of the action $S_{\mathrm{NO}}(\boldsymbol{m}'',t|\boldsymbol{m}_i^*,0)$ along the instanton segments shown in panel (a), plotted parametrically as a function of the $m_z$ coordinate.}
    \label{fig:crf-instantons}
\end{figure}%

The above analytical results provide a benchmark against which we assess the instanton approach. We find that the results obtained in the NO representation agree with the analytical prediction in Eq.~\eqref{eq:crf-barriers-analytic} for both activation barriers. The representative instanton trajectories calculated at the phase transition point are plotted in Fig.~\ref{fig:crf-instantons}(a). In the GC representation, we likewise obtain agreement for the barrier $\mathcal{A}_{u \to \ell}$, whereas the numerical construction of $\ell \to u$ instantons converges only very close to the saddle-node bifurcation at which the fixed points $\ell$ and $s$ emerge. These results confirm the validity of the instanton approach for systems with weak permutation symmetry.

\subsubsection{Manifestations of Hamiltonian nonconvexity} \label{subsubsec:crfnonconvex}
As in the LMG model, the auxiliary Hamiltonians $\mathcal{H}_\text{NO}$ and $\mathcal{H}_\text{GC}$ are nonconvex in the momentum $\boldsymbol{\pi}$, raising the question of whether this nonconvexity is reflected in the behavior of the action~\eqref{eq:action} (with $\boldsymbol{x}=\boldsymbol{m}$). We find the same signature as in the LMG model: the action can transiently decrease, $d_t S_\alpha<0$, during evolution along an instanton trajectory. This behavior is illustrated in Fig.~\ref{fig:crf-instantons}(b) for the  $\alpha=\text{NO}$ representation. Along the $u\to s$ trajectory, the action evolves nonmonotonically, undergoing successive increases and decreases. Along the $\ell\to s$ trajectory, a small decrease is also visible near the end of the trajectory.

\subsubsection{Comparison with SW approach}
\begin{figure}
    \centering
    \includegraphics[width=0.95\linewidth]{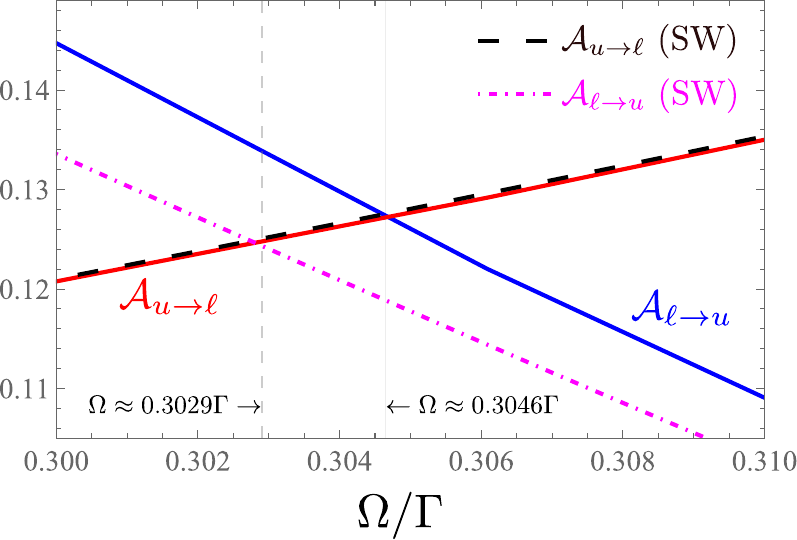}
    \caption{Results for the zero dephasing case $\kappa=0$. Exact activation barriers $\mathcal{A}_{i \rightarrow j}$ calculated from Eq.~\eqref{eq:crf-barriers-analytic} (solid lines) compared with the results for SW approach (dashed lines). Grey solid and dashed vertical lines at $\Omega \approx 0.3046 \Gamma$ and $\Omega \approx 0.3029 \Gamma$ denote the crossing point of $\mathcal{A}_{\ell \rightarrow u}$ and $\mathcal{A}_{u \rightarrow \ell}$ curves for the exact results and SW approach, respectively. Parameters: $\gamma=0.01\Gamma$.
    }
    \label{fig:crf-comparison-sw}
\end{figure}
We further compare the predictions of our approach, which agree with the exact analytical results, with those of the SW approach. The derivation of the corresponding auxiliary Hamiltonian is given in Appendix~\ref{app:auxhamcrf-sw}, with the final expression presented in Eq.~\eqref{eq:auxham-crf-sw}. As in the LMG model, the SW predictions are relatively close to the exact results, so that the differences would be barely discernible on the scale of Fig.~\ref{fig:crf-mzaction-kappa0}(b). They become visible, however, in the magnified view of the vicinity of the phase-transition point shown in Fig.~\ref{fig:crf-comparison-sw}. For the activation barrier $\mathcal{A}_{u \to \ell}$, the exact and SW results are in near-perfect agreement, whereas a noticeable discrepancy appears for $\mathcal{A}_{\ell \to u}$. This discrepancy leads to a small but discernible shift in the predicted position of the first-order phase transition, from $\Omega \approx 0.3046\Gamma$ for the exact result to $\Omega \approx 0.3029\Gamma$ within the SW approach. This corroborates the conclusion of the companion Letter~\cite{ptaszynski2026quantum} that the SW approach does not reproduce the activation barriers exactly. Consequently, the predicted positions of first-order phase transitions are also shifted, although the discrepancy is considerably smaller here than for the model considered in the companion Letter.

\subsection{Finite dephasing ($\kappa \neq 0$) case}
\begin{figure}
    \centering
    \includegraphics[width=0.95\linewidth]{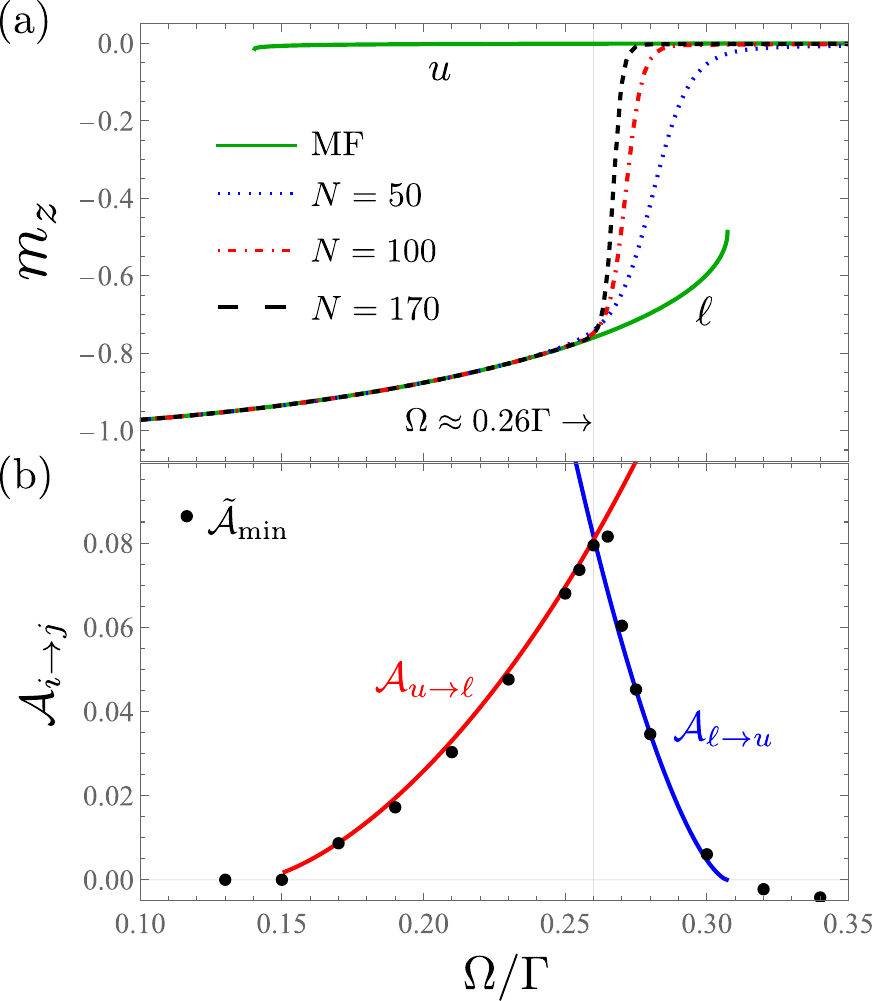}
    \caption{Results for the finite dephasing case $\kappa=0.002\Gamma$. (a) Mean-field (MF) fixed-point magnetization $m_z^*$ (green solid lines) versus finite-size steady-state magnetization $m_{z,\text{ss}}$ obtained using master equation (other lines). (b) Activation barriers $\mathcal{A}_{i \rightarrow j}$ calculated within the NO representation. Grey vertical solid line at $\Omega \approx 0.26 \Gamma$ denotes the crossing point of $\mathcal{A}_{\ell \rightarrow u}$ and $\mathcal{A}_{u \rightarrow \ell}$ curves. Symbols denote the estimator $\tilde{\mathcal{A}}_\text{min}$ defined by Eq.~\eqref{eq:crf-amin}. Parameters: $\gamma=0.01\Gamma$.
    }
    \label{fig:crf-mzaction-kappa0002}
\end{figure}%
We now demonstrate that the instanton approach accurately captures metastability in the presence of finite dephasing, $\kappa \neq 0$, where the exact results discussed above are no longer applicable. Representative results for $\kappa=0.002\Gamma$ are shown in Fig.~\ref{fig:crf-mzaction-kappa0002}. As seen there, the finite-size stationary magnetization behaves similarly to the $\kappa=0$ case presented in Fig.~\ref{fig:crf-mzaction-kappa0}: it remains concentrated around the $\ell$ branch up to a certain value of $\Omega$, beyond which it crosses over toward the $u$ branch. This crossover becomes sharper with increasing $N$. Compared with the $\kappa=0$ case shown in Fig.~\ref{fig:crf-mzaction-kappa0}, the crossover point is shifted from $\Omega \approx 0.3046 \Gamma$ to $\Omega \approx 0.26\Gamma$. Its location is accurately predicted by the crossing of the activation barriers shown in Fig.~\ref{fig:crf-mzaction-kappa0002}(b). These results confirm that the instanton approach can be used to predict the location of first-order phase transitions beyond the exactly solvable regime of $\kappa=0$.

\begin{figure}
    \centering
    \includegraphics[width=0.95\linewidth]{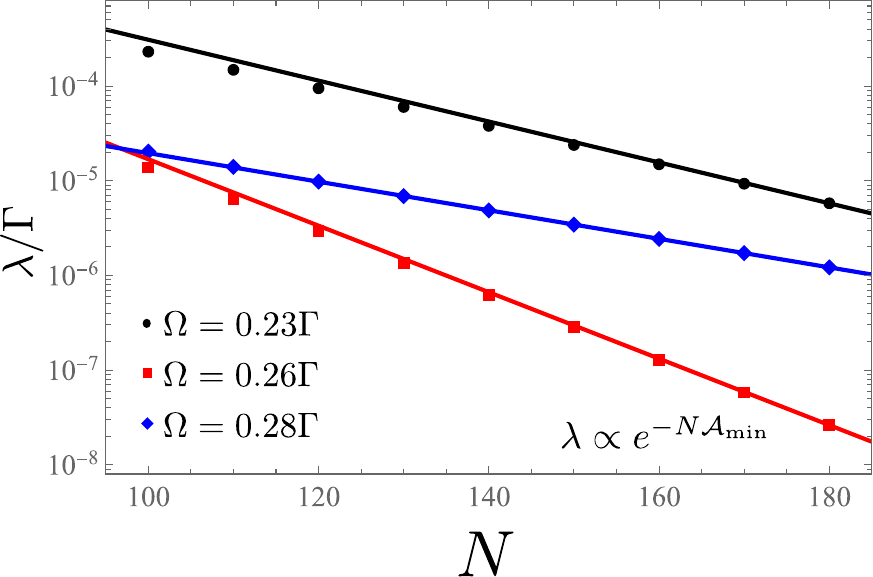}
    \caption{Results for the finite dephasing case $\kappa=0.002\Gamma$. Dots: Liouvillian gap $\lambda$ calculated using QME for selected values of $\Omega$, plotted in the logarithmic scale. Lines: the predicted scaling $\lambda \propto \exp(-N \mathcal{A}_\text{min})$ set to cross the point for $N=180$.
    }
    \label{fig:crf-eigenvalues-kappa0002}
\end{figure}%

The validity of the instanton approach is further supported by its ability to predict the asymptotic exponential scaling of the Liouvillian gap (Fig.~\ref{fig:crf-eigenvalues-kappa0002}). The estimator $\tilde{\mathcal{A}}_\text{min}$ defined in Eq.~\eqref{eq:crf-amin}, plotted in Fig.~\ref{fig:crf-mzaction-kappa0002}(b), also generally agrees well with the smaller of the two activation barriers, $\mathcal{A}_\text{min}$. A small but discernible deviation is observed at $\Omega=0.265\Gamma$, close to the phase-transition point. Its origin is analogous to that discussed in the Supplemental Material of the companion Letter~\cite{ptaszynski2026quantum} and can be attributed to finite-size effects. Although $\mathcal{A}_{u\rightarrow\ell}>\mathcal{A}_{\ell\rightarrow u}$ at this point, the $\ell$ branch remains populated with finite probability for the system sizes $N=170,180$ used to construct the estimator, as indicated by the steady-state magnetization $m_{z,\mathrm{ss}}$ [Fig.~\ref{fig:crf-mzaction-kappa0002}(a)]. Consequently, the Liouvillian gap is determined by the sum of the two switching rates~\cite{carde2026nonperturvative},
\begin{align}
\lambda
\approx \kappa_{\ell\rightarrow u}+\kappa_{u\rightarrow\ell}
=
\omega_{\ell\rightarrow u}e^{-N\mathcal{A}_{\ell\rightarrow u}}
+
\omega_{u\rightarrow\ell}e^{-N\mathcal{A}_{u\rightarrow\ell}}\,,
\end{align}
where $\omega_{i\rightarrow j}$ is a subexponential prefactor satisfying
$\lim_{N\rightarrow\infty}(\ln\omega_{i\rightarrow j})/N=0$.
For finite $N$, both terms therefore contribute to the gap, so that the estimator $\tilde{\mathcal{A}}_\text{min}$ assumes an effective value between $\mathcal{A}_{\ell\rightarrow u}$ and $\mathcal{A}_{u\rightarrow\ell}$, as observed in Fig.~\ref{fig:crf-mzaction-kappa0002}(b).

\section{Conclusions} \label{sec:concl}

We have applied the quantum instanton framework introduced in the companion Letter~\cite{ptaszynski2026quantum} to two microscopically motivated collective spin models with different mechanisms of metastability. The driven-dissipative LMG model preserves total spin angular momentum, whereas local dissipation in the CRF model couples different angular-momentum sectors while retaining weak permutation symmetry. In both cases, the obtained activation barriers identify the first-order transition boundary and predict the exponential scaling of slow system relaxation rates, in agreement with quantum master-equation calculations. For the CRF model without dephasing, the barriers also agree with an available exact result; with finite dephasing, they continue to predict the crossover of the stationary magnetization (a finite-size precursor of a first-order phase transition) and Liouvillian gap scaling beyond that solvable limit. These tests establish the applicability of the approach beyond the minimal, specially symmetric example of the Letter and, in particular, to systems with local dissipative channels.

From a methodological perspective, the instanton approach employs auxiliary Hamiltonians arising from quantum quasiprobability dynamics that are nonconvex in momentum, unlike those of classical stochastic systems. In both models, this nonconvexity manifests as a transient decrease of the action along an instanton and precludes the direct use of minimum-action methods developed for convex Hamiltonians~\cite{grafke2019numerical,gagrani2023action,zakine2023minimum}. We therefore developed a numerical implementation combining collocation and shooting methods. In the systems studied, collocation was more reliable for trajectories escaping nodal attractors, whereas shooting performed better for those escaping spiral attractors, making the two methods complementary.

Our results further show that the choice of quasiprobability representation, and thus the form of the auxiliary Hamiltonian, can strongly affect the numerical construction of instantons even when the physical activation barriers agree. For the LMG model, the Husimi (H) and P representations yield different instanton trajectories but consistent barriers whenever both calculations converge. The H representation gave converged results throughout the parameter range explored, whereas convergence in the P representation was more limited. For CRF, the normal-order (NO) representation provided both barriers across the studied regimes and reproduced the exact zero-dephasing benchmark; the generalized coherent-state (GC) representation agreed for the $u\to\ell$ barrier, but converged for the $\ell\to u$ instanton only near the $\ell$--$s$ saddle-node bifurcation. We note that numerical convergence has also been reported as an issue in the Keldysh path-integral treatment of activation barriers in a bistable bosonic oscillator~\cite{lee2024real}. This raises the question of how, in bosonic systems, our quasiprobability-based approach compares with the Keldysh formulation in terms of numerical convergence. Systematic comparisons of representations and different boundary-value algorithms could identify choices that make instanton calculations more robust and clarify how numerical tractability depends on the representation and on the solver.

Finally, comparison of our framework with the semiclassical Wigner (SW) approach reveals that its accuracy depends strongly on the model and parameter regime. By truncating the quasiprobability dynamics at second order in derivatives, the SW approach yields an effectively classical Fokker--Planck problem amenable to minimum-action methods~\cite{xiang2025switching}, but omits contributions that can change activation barriers at leading exponential order. In the LMG regimes studied here, the SW approach accurately predicts stationary magnetization and Liouvillian-gap scaling, although $\mathcal{A}_{\ell \to u_\pm}$ barrier differs appreciably when $\mathcal{A}_{\ell \to u_\pm}$ and $\mathcal{A}_{u_\pm \to \ell}$ barriers are strongly asymmetric. For the CRF model without dephasing, a discrepancy in the barriers produces a small but discernible shift in the predicted first-order transition point. Together with the much larger deviations found for the model of the companion Letter~\cite{ptaszynski2026quantum} and for bosonic systems~\cite{drummond1989quantum,kinsler1991quantum}, these results show that the SW approach cannot be assumed accurate in general, even though it works well for some models and parameter ranges. Identifying conditions under which the SW approach reliably describes observable metastable behavior, or instead misses significant non-Gaussian effects, remains an important open problem.

\begin{acknowledgments}
M.C.\ and K.P.\ acknowledge the financial support of the National Science Centre, Poland, under the project No.\ 2023/51/D/ST3/01203, and M.E.\ of the Fond National de la Recherche-FNR, Luxembourg, CORE project NEQPHASETRANS (C24/MS/18933049).
\end{acknowledgments}

\section*{Data availability}
The data and source code that support the findings of this article are openly
available at~\cite{zenodo,github}.

\appendix

\section{Mean-field dynamics} \label{app:meanfield}
Here, we sketch the derivation of the mean-field equations~\eqref{eq:meanfieldising} and~\eqref{eq:meanfieldcrf}, which is rigorously proven in Refs.~\cite{alicki1983nonlinear,benatti2016non,benatti2018quantum,fiorelli2023mean}. The starting point is the Heisenberg evolution of spin operators
\begin{align} \label{eq:heis}
d_t \hat{J}_k=\mathcal{L}^\dagger \hat{J}_k \,,
\end{align}
where $\mathcal{L}^\dagger$ is the superoperator adjoint to $\mathcal{L}$. It is obtained from $\mathcal{L}$ by replacement $\hat{H} \rightarrow -\hat{H}$ and $\mathcal{D}[\hat{A}] \bullet \rightarrow \mathcal{D}^\dagger[\hat{A}] \bullet\equiv\hat{A}^\dagger \bullet \hat{A}-\{\hat{A}^\dagger \hat{A},\bullet\}/2$. From the corresponding equations of motion for the spin operators, the mean-field dynamics is obtained by replacing operators with classical variables $\hat{J}_\pm \rightarrow J(m_x \pm i m_y)$ and $\hat{J}_{x,y,z} \rightarrow J m_{x,y,z}$, dividing both sides by $J$, and omitting terms of order $O(1/J)$.

\section{Local approximation of stable and unstable manifolds} \label{app:locapp}
\subsection{Linear approximation}
As discussed in Sec.~\ref{sec:framework}, an $i\to j$ instanton comprises a $\boldsymbol{\pi}\neq\boldsymbol{0}$ segment from $i$ to an intermediate unstable mean-field fixed point $\nu$, followed by a $\boldsymbol{\pi}=\boldsymbol{0}$ relaxation segment from $\nu$ to $j$. Both are heteroclinic connections between fixed points of Hamilton's equations~\eqref{eq:hamilton-eq}~\cite{ptaszynski2026quantum}. We focus on the first segment, which lies in the unstable manifold of $(\boldsymbol{x}_i^*,\boldsymbol{0})$ and the stable manifold of $(\boldsymbol{x}_\nu^*,\boldsymbol{0})$. The collocation method approximates these asymptotic conditions at both ends of this segment; the shooting method uses a local approximation of the unstable manifold at its starting point.

Near either fixed point, the linear approximation is determined by the Jacobian of Hamilton's equations~\eqref{eq:hamilton-eq}. With row-vector convention $\boldsymbol{q}=(\boldsymbol{x}-\boldsymbol{x}^*,\boldsymbol{\pi})$ and $\dot{\boldsymbol{q}}=\boldsymbol{q}\mathbb{K}+O(\Vert\boldsymbol{q}\Vert_2^2)$, it reads
\begin{align} \label{eq:conjjac}
\mathbb{K} \equiv [\partial_{q_k} \dot{q}_l]_{k,l} \, \big\vert_{\boldsymbol{q}=\boldsymbol{0}}
=\begin{pmatrix} \mathbb{J} & 0 \\ 2\mathbb{D} & -\mathbb{J}^\intercal \end{pmatrix} \,,
\end{align}
where $\mathbb{J}$ is the mean-field Jacobian~\eqref{eq:mfjacobian} and
\begin{align}
\mathbb{D} \equiv \frac{1}{2} \left[\partial_{\pi_k} \partial_{\pi_l} \mathcal{H}_\alpha \right]_{k,l}  \big \vert_{\boldsymbol{q}=\boldsymbol{0}}
\end{align}
is the diffusion matrix. Provided the eigenvalues of $\mathbb{J}$ have nonzero real parts, the Hamiltonian fixed point is hyperbolic and has $d$ stable and $d$ unstable directions. We denote the corresponding left eigenvectors of $\mathbb{K}$ by $\boldsymbol{j}_m^s$ and $\boldsymbol{j}_m^u$ ($m=1,\ldots,d$); the resulting real eigenspaces are tangent to the local stable and unstable manifolds, respectively. For a complex-conjugate pair, the real and imaginary parts of either eigenvector span the corresponding real subspace.

We use this local approximation to choose the initial points for the shooting method. For escape from a spiral attractor with $d=2$, the two unstable eigenvectors form a complex-conjugate pair, $\boldsymbol{j}^u_2=(\boldsymbol{j}^u_1)^*$. We parameterize the real initial displacement by a single phase $\varphi$,
\begin{align}
\boldsymbol{q}_i(0)
= \frac{\varepsilon}{2} \left(e^{i\varphi}\boldsymbol{j}^u_1
+e^{-i\varphi}\boldsymbol{j}^u_2\right) \,,
\end{align}
with chosen small amplitude $\varepsilon$, $\boldsymbol{q}_i \equiv (\boldsymbol{x}-\boldsymbol{x}_i^*,\boldsymbol{\pi})$, and the normalization $\boldsymbol{j}_{1,2}^u \cdot (\boldsymbol{j}_{1,2}^u)^*=1$.

\subsection{A refined local approximation}

The linear approximation was also used in Ref.~\cite{lee2024real} to define the boundary conditions for the collocation method. In our implementation, we found that a refinement retaining the nonlinear mean-field drift improves convergence. We first write a vector in the tangent space to the stable ($s$) or unstable ($u$) manifold as
\begin{align}
\boldsymbol{q}^{s,u} = \sum_{m=1}^d \zeta_m \boldsymbol{j}_m^{s,u} =\boldsymbol{\zeta} \mathbb{S}^{s,u} \,,
\end{align}
where $\boldsymbol{\zeta} \equiv (\zeta_1,\ldots,\zeta_d)$ contains the amplitudes (chosen so that $\boldsymbol{q}^{s,u}$ is real) and the eigenvectors are arranged as the rows of
\begin{align}
\mathbb{S}^{s,u}_{d \times 2d} \equiv \begin{pmatrix} \boldsymbol{j}_1^{s,u} \\  \vdots \\ \boldsymbol{j}_d^{s,u} \end{pmatrix} = \begin{pmatrix} \mathbb{X}^{s,u}_{d \times d} & \mathbb{P}^{s,u}_{d \times d}  \end{pmatrix} \,.
\end{align}
Writing  $\boldsymbol{r} \equiv \boldsymbol{x}-\boldsymbol{x}^*$, we have
\begin{align}
\boldsymbol{r}=\boldsymbol{\zeta} \mathbb{X} \,, \quad \boldsymbol{\pi}=\boldsymbol{\zeta} \mathbb{P} \,,
\end{align}
where the superscript $s$ or $u$ has been suppressed. If $\mathbb{X}$ is invertible, the tangent-space relation becomes
\begin{align} \label{eq:tansprel}
\boldsymbol{\pi}=\boldsymbol{r} \mathbb{X}^{-1} \mathbb{P} \,.
\end{align}
To refine this tangent-space relation at finite displacement $\boldsymbol{r}$, we rescale the momentum while retaining its direction:
\begin{align} \label{eq:locaappansatz}
\boldsymbol{\pi}=\beta(\boldsymbol{r})\boldsymbol{r} \mathbb{X}^{-1} \mathbb{P} \,.
\end{align}
Since the instanton satisfies $\mathcal{H}_\alpha=0$~\cite{ptaszynski2026quantum}, we determine the scalar $\beta(\boldsymbol{r})$ by requiring a local approximation to the Hamiltonian to vanish. This approximation retains the full nonlinear mean-field drift $d_t \boldsymbol{x}=\boldsymbol{h}(\boldsymbol{x})=\nabla_{\boldsymbol{\pi}} \mathcal{H}_\alpha(\boldsymbol{x},\boldsymbol{\pi})\, \vert_{\boldsymbol{\pi}=\boldsymbol{0}}$ but evaluates the momentum-quadratic term at the fixed point:
\begin{align} \label{eq:quadapproxham}
\mathcal{H}_\alpha \approx \boldsymbol{h}(\boldsymbol{x}^*+\boldsymbol{r}) \boldsymbol{\pi}^\intercal+\boldsymbol{\pi} \mathbb{D} \boldsymbol{\pi}^\intercal \,.
\end{align}
Requiring the right-hand side of Eq.~\eqref{eq:quadapproxham} to vanish and substituting Eq.~\eqref{eq:locaappansatz}, we find, whenever the denominator is nonzero,
\begin{align}
\beta(\boldsymbol{r})=-\frac{\boldsymbol{h} (\boldsymbol{x}^*+\boldsymbol{r}) \left(\mathbb{X}^{-1} \mathbb{P} \right)^\intercal \boldsymbol{r}^\intercal }{\boldsymbol{r} \mathbb{X}^{-1} \mathbb{P} \mathbb{D} \left(\mathbb{X}^{-1} \mathbb{P} \right)^\intercal \boldsymbol{r}^\intercal} \,.
\end{align}

In collocation, Eq.~\eqref{eq:locaappansatz} supplies $d$ scalar conditions at the initial point, near the unstable manifold of $(\boldsymbol{x}_i^*,\boldsymbol{0})$, and another $d$ at the final point, near the stable manifold of $(\boldsymbol{x}_\nu^*,\boldsymbol{0})$. These are the $2d$ manifold boundary conditions specified in Sec.~\ref{subsubsec:collocation}.

\section{Initial guesses for collocation and comparison with shooting} \label{app:initcol}

The collocation solver of Sec.~\ref{subsubsec:collocation} requires initial guesses for both the trajectory and its duration, and its convergence depends strongly on their quality. For escape from a nodal attractor in an effectively two-dimensional ($d=2$) state space, we construct these guesses from the slow eigendirections near the initial attractor and the target saddle. Let the real, negative eigenvalues of the mean-field Jacobian~\eqref{eq:mfjacobian} at the attractor be ordered as $0>a_1>a_2$. The phase-space Jacobian~\eqref{eq:conjjac} then has eigenvalues $\{a_1,a_2,-a_1,-a_2\}$; its two unstable eigenvalues are $|a_1|<|a_2|$. To see why the slow eigendirection provides a useful initial tangent, consider the instanton trajectory backward in time as it approaches the attractor. In the linearized dynamics, its phase-space displacement along the unstable manifold is
\begin{align} \label{eq:initdisplacementdyn}
\boldsymbol{q}_i(t)\simeq
 \zeta_1 e^{|a_1|t}\boldsymbol{j}_1^u
+\zeta_2 e^{|a_2|t}\boldsymbol{j}_2^u \,,\qquad t\to-\infty \,.
\end{align}
If $\zeta_1\neq 0$, the fast unstable component along $\boldsymbol{j}_2^u$ is exponentially suppressed relative to the slow component along $\boldsymbol{j}_1^u$ as $t\to-\infty$: the ratio of their amplitudes scales as $|\zeta_2/\zeta_1|e^{(|a_2|-|a_1|)t}$. We therefore choose $\boldsymbol{j}_1^u\equiv\boldsymbol{j}_\text{slow}^u$ as the initial tangent of the trial path, particularly when $|a_2|\gg|a_1|$.

For the intermediate saddle $\nu$ in the models studied here, let the mean-field eigenvalues be $b_+>0>b_-$. Its phase-space Jacobian~\eqref{eq:conjjac} has stable eigenvalues $b_-$ and $-b_+$. Provided their magnitudes differ and the corresponding component is nonzero, a trajectory approaching the saddle is asymptotically tangent to the slow stable eigenvector $\boldsymbol{k}_\text{slow}^s$, whose eigenvalue is $-\min\{b_+,|b_-|\}$. We use this vector for the tangent at the other end of the trial path. 

We then define a four-dimensional cubic Hermite spline $\boldsymbol{f}(u)=(\boldsymbol{x}(u),\boldsymbol{\pi}(u))$, $u\in[0,1]$, between the phase-space fixed points $\boldsymbol{p}_i=(\boldsymbol{x}_i^*,\boldsymbol{0})$ and $\boldsymbol{p}_\nu=(\boldsymbol{x}_\nu^*,\boldsymbol{0})$:
\begin{align}
\boldsymbol{f}(u)=
 h_{00}(u)\boldsymbol{p}_i+h_{10}(u)\boldsymbol{v}_i
+h_{01}(u)\boldsymbol{p}_\nu+h_{11}(u)\boldsymbol{v}_\nu \,,
\end{align}
where $h_{00}=2u^3-3u^2+1$, $h_{10}=u^3-2u^2+u$, $h_{01}=-2u^3+3u^2$, and $h_{11}=u^3-u^2$. Writing $\Delta=\Vert\boldsymbol{x}_\nu^*-\boldsymbol{x}_i^*\Vert_2$, we choose the tangent vectors
\begin{align}
\boldsymbol{v}_i=y_i\Delta\boldsymbol{j}_\text{slow}^u \,,
\qquad
\boldsymbol{v}_\nu=y_\nu\Delta\boldsymbol{k}_\text{slow}^s \,,
\end{align}
with $y_i,y_\nu\in\{-1,1\}$ and both eigenvectors normalized to unit Euclidean norm. Thus $\boldsymbol{f}'(0)=\boldsymbol{v}_i$ and $\boldsymbol{f}'(1)=\boldsymbol{v}_\nu$. The signs of $y_i,y_\nu$ select the branches of the local eigendirections; we test their four combinations when constructing the initial guess. The choice of tangent length $\Delta$ is heuristic and can be adjusted to improve convergence.

We then restrict the curve to $u\in[u_0,u_1]$, choosing its endpoints at a state-space distance $\epsilon$ from their respective fixed points: $\Vert\boldsymbol{x}(u_0)-\boldsymbol{x}_i^*\Vert_2=\Vert\boldsymbol{x}(u_1)-\boldsymbol{x}_\nu^*\Vert_2=\epsilon$. The collocation boundary conditions fix the initial distance $\epsilon$; the same distance at the final point is used only to construct the guess. To estimate its duration, we divide the phase-space arc-length element by the speed of Hamiltonian flow along the trial curve. With $\boldsymbol{F}_\alpha(\boldsymbol{x},\boldsymbol{\pi})=(\nabla_{\boldsymbol{\pi}}\mathcal{H}_\alpha,-\nabla_{\boldsymbol{x}}\mathcal{H}_\alpha)$, this gives
\begin{align}
T_{\mathrm{guess}}=
\int_{u_0}^{u_1}
\frac{\Vert\boldsymbol{f}'(u)\Vert_2}
{\Vert\boldsymbol{F}_\alpha[\boldsymbol{f}(u)]\Vert_2}\,du \,.
\end{align}
The numerator makes this estimate independent of the parametrization of the trial curve. Both $\boldsymbol{f}(u)$ and $T_{\mathrm{guess}}$ are supplied to the collocation solver, which then determines the finite-time trajectory and its duration.

This construction works best when the nodal attractor has well-separated rates, $|a_1|\ll|a_2|$, so that the slow unstable eigendirection provides a clear initial tangent. It becomes less effective as $|a_2|$ approaches $|a_1|$. For spiral attractors, there is no distinguished real slow eigendirection, and we use the shooting method instead. Conversely, in the strongly separated regime where collocation works best, shooting is highly sensitive to the initial direction: as implied by Eq.~\eqref{eq:initdisplacementdyn}, a small component along $\boldsymbol{j}_2^u$ grows relative to the slow component by a factor proportional to $e^{(|a_2|-|a_1|)t}$ during forward propagation. Numerical precision then limits how accurately the optimal initial condition can be found. This explains why collocation and shooting methods performed best in complementary regimes in our calculations.

\section{Master equation for the LMG model} \label{app:masteq-lmg}
To numerically determine the steady state and Liouvillian eigenvalues for the LMG model, we express the master equation in the Liouville space representation~\cite{machnes2014surprising, amshallem2015approaches, uzdin2016speed,minganti2018spectral}. The $(2J+1) \times (2J+1)$ density matrix $\hat{\rho}$, representing the system state for a fixed value of $J$, is transformed into the column vector $|\rho \rrangle$ 
obtained by stacking the matrix columns. The steady state is given by the stationary solution of the master equation,
\begin{align} \label{eq:masteqvectorized}
d_t |\rho \rrangle = \bar{\mathcal{L}} |\rho \rrangle=0  \,,
\end{align}
where 
\begin{align} \label{eq:lmgliov}
	&\bar{\mathcal{L}} =-i \left( \mathds{1}_{2J+1} \otimes \bar{H}_J-\bar{H}_J^T \otimes \mathds{1}_{2J+1} \right) \\ \nonumber &+\frac{\Gamma}{J} \left[\bar{J}_-^* \otimes \bar{J}_- - \frac{1}{2}\mathds{1}_{2J+1} \otimes \bar{J}_-^\dagger \bar{J}_- - \frac{1}{2} \left(\bar{J}_-^\dagger \bar{J}_- \right)^T \otimes \mathds{1}_{2J+1} \right] \,,
\end{align}
is the matrix representing the Liouvillian $\mathcal{L}$ for a fixed value of $J$. Here,
 $^*$ denotes the complex conjugate, $^\dagger$ denotes the Hermitian conjugate, $\bar{H}_J=\omega \bar{J}_z-\chi \bar{J}_x^2/J$, while $\bar{J}_k$ are the matrix representations of $\hat{J}_k$ operators for a fixed value of $J$.

We further note that the system Liouvillian possesses a discrete $C_2$ symmetry generated by the parity superoperator $\mathcal{P}\hat{\rho}=\Pi\hat{\rho}\Pi^\dagger$, with $\Pi=e^{i\pi\hat{J}_z}$. This symmetry originates from the structure of both the coherent and dissipative parts of the dynamics: the Hamiltonian contains only $\hat{J}_z$ and $\hat{J}_x^2$ operators, which are invariant under a $\pi$ rotation about the $z$ axis, whereas the jump operator transforms as $\Pi\hat{J}_-\Pi^\dagger=-\hat{J}_-$. Since the dissipative terms are bilinear in the jump operators, this sign change cancels, and the full Liouvillian remains invariant under the rotation. The resulting $C_2$ symmetry is isomorphic to the $\mathbb{Z}_2$ symmetry used in the Liouvillian classification of Ref.~\cite{minganti2023dissipative}, so that the two descriptions represent the same underlying symmetry structure. Accordingly, following the general $\mathbb{Z}_n$ symmetry decomposition discussed in Ref.~\cite{minganti2023dissipative}, the matrix $\bar{\mathcal{L}}$ can be brought, by simultaneous permutations of rows and columns, to a block-diagonal form with independent even- and odd-parity blocks, allowing the two blocks to be treated separately. This substantially reduces the computational cost and memory requirements of the eigenvalue and steady-state calculations. The Liouvillian eigenvalues $\lambda_i$ associated with even or odd eigenstates $\hat{\rho}_i$ are then obtained from the corresponding parity block, a property used in Sec.~\ref{sec:liovgap-lmg} to define the symmetric Liouvillian gap $\lambda_\text{sym}$.

Numerically, the steady state satisfying~\eqref{eq:masteqvectorized}, Liouvillian gap, and symmetric Liouvillian gap for the LMG model were determined via partial sparse eigendecomposition using SciPy~\cite{2020SciPy-NMeth} and SPLU decomposition provided by UMFPACK~\cite{davis2004algorithm}. 
The stationary magnetization values were calculated as
\begin{align}
  & m_{k,\text{ss}} =\frac{1}{J} \llangle J_k|\rho \rrangle \,,
  \end{align}
where the row vector $\llangle J_k|$ is constructed by stacking the rows of the matrix $\bar{J}_k$.

\section{Derivation of the auxiliary Hamiltonians for the LMG model} \label{app:auxhamlmg}
\subsection{Derivations for H and P representations}  \label{app:auxhamlmg-derhp}
Here we present the method for deriving the differential operators $L_\alpha(\boldsymbol{x},\nabla_{\boldsymbol{x}})$, and hence the auxiliary Hamiltonians $\mathcal{H}_\alpha(\boldsymbol{x},\boldsymbol{\pi})$, for the LMG model. Although this discussion is largely identical to that presented in the companion Letter~\cite{ptaszynski2026quantum} due to shared strong permutation symmetry, we repeat it here to make the present article self-contained. First, for H and P representations, we represent the action of the spin operators on $|\boldsymbol{x}\rangle \langle \boldsymbol{x}|$, where $|\boldsymbol{x}\rangle$ is the coherent spin state expressed in the stereographic coordinates, by the corresponding differential operators~\cite{narducci1975multitime,altland2012quantum,mandt2015stochastic,burkle2020probabilistic}:
\begin{align}
\hat{A} |\boldsymbol{x}\rangle \langle \boldsymbol{x}| \equiv \mathcal{J}[\hat{A}]|\boldsymbol{x}\rangle \langle \boldsymbol{x}| \,, \quad |\boldsymbol{x}\rangle \langle \boldsymbol{x}| \hat{A} \equiv \mathcal{J}^*[\hat{A}^\dagger]|\boldsymbol{x}\rangle \langle \boldsymbol{x}| \,,
\end{align}
where
\begin{align} \nonumber
\mathcal{J}[\hat{J}_+]&=\frac{1}{2} \left[ \frac{\partial}{\partial v}+i\frac{\partial}{\partial w}+\frac{4J(v+iw)}{v^2+w^2+1} \right] \,, \\ \nonumber
\mathcal{J}[\hat{J}_-]&=\frac{1}{2} \left[ -(v-iw)^2\left(\frac{\partial}{\partial v}+i\frac{\partial}{\partial w}\right)+\frac{4J(v-iw)}{v^2+w^2+1} \right] \,, \\ \label{eq:corrules}
\mathcal{J}[\hat{J}_z]&=\frac{1}{2} \left[(v-iw)\left(\frac{\partial}{\partial v}+i\frac{\partial}{\partial w}\right)+\frac{2J(v^2+w^2-1)}{v^2+w^2+1} \right] \,.
\end{align}
Here we use the operators from Ref.~\cite{burkle2020probabilistic} expressed in stereographic coordinates. The operators $L_\alpha(\boldsymbol{x},\nabla_{\boldsymbol{x}})$ are given by the correspondence rules~\cite{burkle2020probabilistic,altland2012quantum,mandt2015stochastic}
\begin{subequations}
\begin{align}
L_\text{H}(\boldsymbol{x},\nabla_{\boldsymbol{x}}) |\boldsymbol{x}\rangle \langle \boldsymbol{x}| &\equiv \mathcal{L}^\dagger |\boldsymbol{x}\rangle \langle \boldsymbol{x}| \,, \\
L_\text{P}^\dagger(\boldsymbol{x},\nabla_{\boldsymbol{x}}) |\boldsymbol{x}\rangle \langle \boldsymbol{x}| &\equiv \mathcal{L}|\boldsymbol{x}\rangle \langle \boldsymbol{x}| \,.
\end{align}
\end{subequations}
Here, $\mathcal{L}^\dagger$ is defined below Eq.~\eqref{eq:heis} and $L_\text{P}^\dagger$ is the adjoint to $L_\text{P}$. Expanding $L_\text{P}^\dagger$ as
\begin{align}
L_\text{P}^\dagger(\boldsymbol{x},\nabla_{\boldsymbol{x}})=\sum_{a,b} L_\text{P}^{(a,b)} (\boldsymbol{x}) \frac{\partial^a}{\partial v^a} \frac{\partial^b}{\partial w^b} \,,
\end{align}
the operator $L_\text{P}$ can be obtained as
\begin{align}
L_\text{P}(\boldsymbol{x},\nabla_{\boldsymbol{x}})=\sum_{a,b} (-1)^{a+b}\frac{\partial^a}{\partial v^a} \frac{\partial^b}{\partial w^b} L_\text{P}^{(a,b)} (\boldsymbol{x}) \,.
\end{align}
Employing Eq.~\eqref{eq:auxhamdef} with $V=J$, the Hamiltonian $\mathcal{H}_\text{P}$ can also be obtained directly from $L_\text{P}^\dagger$ as $\mathcal{H}_\text{P}(\boldsymbol{x},\boldsymbol{\pi})=\lim_{J \rightarrow \infty} J^{-1} L_\text{P}^\dagger(\boldsymbol{x},J \boldsymbol{\pi})$. In general, to derive $\mathcal{H}_\alpha$, it is simples to make substitution $\partial_{v,w} \rightarrow - J \pi_{v,w}$ ($\alpha=\text{H}$) or $\partial_{v,w} \rightarrow J \pi_{v,w}$ ($\alpha=\text{P}$) already at the level of operators~\eqref{eq:corrules}, so one does not have to take care with operator ordering.

\subsection{Derivation for SW approach} \label{app:auxhamlmg-dersw}
The derivations for the SW approach follow a method described in Ref.~\cite{dubois2021semi}. The Lindblad equation~\eqref{eq:masteqising}
is represented by the differential operator $L_\text{SW}(\boldsymbol{x},\nabla_{\boldsymbol{x}})$ that describes the evolution of the Wigner distribution $p_\text{W}(\boldsymbol{x})$, which is truncated after the second derivative term. The operator $L_\text{SW}$ has Fokker--Planck form 
\begin{align}
L_\text{SW}(\boldsymbol{x},\nabla_{\boldsymbol{x}})=-\sum\limits_{\mathclap{k \in \{v,w\}}} \partial_k h_k(\boldsymbol{x}) + J^{-1} \sum\limits_{\mathclap{k,l \in \{v,w\}}} \partial_k \partial_l D_{kl}^\text{SW}(\boldsymbol{x}) \,, 
\end{align}
with drift and diffusion terms
\begin{align} \nonumber
&h_k(\boldsymbol{x})=\{k,H \}_\text{PB} \\ \nonumber &-\sum_j \frac{i \Gamma}{2J} \left(J_- \left\{k,J_-^* \right\}_\text{PB}+J_-^* \left\{J_- ,k\right\}_\text{PB} \right)+O(1/J)\,, \\
&D_{kl}^\text{SW}(\boldsymbol{x})=\sum_{j} \frac{\Gamma}{4} \big( \{k,J_-\}_\text{PB} \{l,J_-^*\}_\text{PB}+\text{c.c} \big) \,.
\end{align}
Here, the functions $H$, $J_-$ are obtained from $\hat{H}$, $\hat{J}_-$ by replacing the operators $\hat{J}_k$ with scalars $J_k$. The object
\begin{align}
\{F,G\}_\text{PB} \equiv \boldsymbol{J} \cdot \left( \nabla_{\boldsymbol{J}} F \times  \nabla_{\boldsymbol{J}} G \right) \,,
\end{align}
where $\boldsymbol{J}=(J_x,J_y,J_z)$, is the spin Poisson bracket. For the stereographic coordinates $\boldsymbol{x} \equiv (v,w)$ considered, we further have $v=J_x/(J-J_z)$ and $w=J_y/(J-J_z)$.

\subsection{Explicit form of the auxiliary Hamiltonians} \label{app:auxhamlmg-res}
Following these derivations, for all representations, the auxiliary Hamiltonians take the quadratic form
\begin{align}
\mathcal{H}_{\alpha}(v,w)=\sum\limits_{k \in \{v,w\}} h_k \pi_k + \sum_{k,l \in \{v,w\}} D_{kl}^\alpha \pi_k \pi_l \,,
\end{align}
with the drift terms $h_v$, $h_w$ given by Eq.~\eqref{eq:lmgmeanfieldster} 
and the diffusion terms
\begin{subequations}
\begin{align} \nonumber
&D_{vv}^\alpha=\frac{\Gamma}{8}  \left[v^2+(w-1)^2\right] \left[v^2+(w+1)^2\right] \\
&+ \frac{A_\alpha}{8} \left\{4 \chi v w \left(v^2-w^2-1\right)+\Gamma  \left[\left(v^2+w^2\right)^2-1\right]\right\} \,,\\ \nonumber
&D_{ww}^\alpha=\frac{\Gamma}{8} \left[(v-1)^2+w^2\right] \left[(v+1)^2+w^2\right] \\
&+\frac{A_\alpha}{8} \left\{4 \chi v w \left(w^2-v^2+1\right)+\Gamma  \left[\left(v^2+w^2\right)^2-1\right]\right\} \,, \\ \nonumber
&D^\alpha_{vw}=D^\alpha_{wv}=\frac{\Gamma  v w}{2} \\ &-\frac{A_\alpha \chi}{8} \left[v^4-2 v^2 \left(3 w^2+1\right)+\left(w^2+1\right)^2\right] \,,
\end{align}
\end{subequations}
where $A_\alpha=1,0,-1$ for P, SW, and H representations, respectively. The diffusion coefficients obey a known property $D_{kl}^\text{SW}=(D_{kl}^\text{H}+D_{kl}^\text{P})/2$~\cite{carmichael1999statistical1}. We note that the Hamiltonians $\mathcal{H}_\alpha$ are not convex for $\alpha \in \{\text{H},\text{P}\}$ representations because the corresponding diffusion matrices $\mathbb{D}^\alpha \equiv [D_{kl}^\alpha]_{kl}$ are not positive definite.

\section{Master equation for the CRF model} \label{app:master-crf}
A numerically tractable form of the master equation for the CRF model is obtained using  the approach from Ref.~\cite{shammah2018open}. This method employs the fact that the permutation invariant state of $N$ spins-1/2 can be expressed as
\begin{align}
\hat{\rho}=\bigoplus_J \left( \rho_J \otimes \mathds{1}_{\text{dim}_J}/\text{dim}_J \right) \,,
\end{align}
where $J \in \{N/2 \mod 1,\ldots,N/2-1,N/2 \}$ are the eigenvalues of the total angular momentum, and 
\begin{align}
\text{dim}_J=(2J+1)\frac{N!}{\left( \tfrac{N}{2}+J+1 \right)!\left( \tfrac{N}{2}-J\right)!}
\end{align}
is the dimension of the permutation-invariant subspace with given $J$. The elements of $(2J+1) \times (2J+1)$ matrices $\rho_J$ are denoted as $\rho_{J,J_z,J_z'}$ with $J_z,J_z' \in \{ -J,-J+1,\ldots,J \}$. The density matrix can then be effectively represented by the column vector $|\rho \rrangle$ being the union of the vectors $|\rho_J \rrangle$, each 
obtained by stacking the columns of the matrices $\rho_J$. The steady state is given by the stationary solution of the master equation,
\begin{align} \label{eq:masteqvectorized-crf}
d_t |\rho \rrangle = \bar{\mathcal{L}} |\rho \rrangle=0  \,,
\end{align}
where $\bar{\mathcal{L}} = \bigoplus_J \bar{\mathcal{L}}_J+\bar{\mathcal{L}}_\text{loc}$. Here, the matrices $\bar{\mathcal{L}}_J$ describe the global evolution that preserves the total angular momentum $J$. They are defined similarly to Eq.~\eqref{eq:lmgliov},
\begin{align} 
	&\bar{\mathcal{L}}_J =-i \left( \mathds{1}_{2J+1} \otimes \bar{H}_J-\bar{H}_J^T \otimes \mathds{1}_{2J+1} \right) \\ \nonumber &+\frac{\Gamma}{N} \left[\bar{J}_-^* \otimes \bar{J}_- - \frac{1}{2}\mathds{1}_{2J+1} \otimes \bar{J}_-^\dagger \bar{J}_- - \frac{1}{2} \left(\bar{J}_-^\dagger \bar{J}_- \right)^T \otimes \mathds{1}_{2J+1} \right] \,,
\end{align}
with $\bar{H}_J=\Omega \bar{J}_x$. The matrix $\bar{\mathcal{L}}_\text{loc}$ describes the local dissipation and dephasing. It is defined such that the generated evolution $d_t |\rho \rrangle=\bar{\mathcal{L}}_\text{loc} |\rho \rrangle$ corresponds to the equations of motion
\begin{align} \nonumber
&d_t \rho_{J,J_z,J_z'} = \sum_{k=-1}^1 R^{(k)}_{J+k,J_z+1,J_z'+1} \rho_{J+k,J_z+1,J_z'+1} \\ &+\sum_{k=-1}^1 K^{(k)}_{J+k,J_z,J_z'} \rho_{J+k,J_z,J_z'}  -\frac{\gamma}{2} (N+J_z+J_z') \rho_{J,J_z,J_z'} \,,
\end{align}
where
\begin{subequations}
\begin{align} 
R^{(1)}_{J,J_z,J_z'} &= \frac{\gamma}{2} B_-^{J,J_z} B_-^{J,J_z'} \frac{\tfrac{N}{2}+J+1}{J(2J+1)} \,, \\
R^{(0)}_{J,J_z,J_z'} &= \frac{\gamma}{2} A_-^{J,J_z} A_-^{J,J_z'} \frac{\tfrac{N}{2}+1}{J(J+1)} \,, \\ 
R^{(-1)}_{J,J_z,J_z'} &= \frac{\gamma}{2} D_-^{J,J_z} D_-^{J,J_z'} \frac{\tfrac{N}{2}-J}{(J+1)(2J+1)} \,, \\
K^{(1)}_{J,J_z,J_z'} &= \kappa B_z^{J,J_z} B_z^{J,J_z'} \frac{\tfrac{N}{2}+J+1}{J(2J+1)} \,, \\
K^{(0)}_{J,J_z,J_z'} &=\kappa \left[ J_z J_z'\frac{\tfrac{N}{2}+1}{J(J+1)} - \frac{N}{2} \right] \,, \\ 
K^{(-1)}_{J,J_z,J_z'} &= \kappa D_z^{J,J_z} D_z^{J,J_z'} \frac{\tfrac{N}{2}-J}{(J+1)(2J+1)} \,,
\end{align}
\end{subequations}
with 
\begin{subequations}
\begin{align}
&A_-^{J,J_z}=\sqrt{(J + J_z)(J - J_z+1)} \,, \\
&B_-^{J,J_z}=-\sqrt{(J+J_z)(J+J_z-1)} \,, \\ &D_-^{J,J_z}=\sqrt{(J-J_z+1)(J-J_z+2)} \,, \\
&B_z^{J,J_z}=\sqrt{(J+J_z)(J-J_z)} \,, \\ 
&D_z^{J,J_z}=\sqrt{(J+J_z+1)(J-J_z+1)} \,. 
\end{align}
\end{subequations}

Numerically, the Liouvillian gap and the steady state satisfying equation~\eqref{eq:masteqvectorized-crf} for the CRF model were determined via partial eigendecomposition using SLEPc’s Krylov–Schur eigensolver~\cite{Hernandez:2005:SSF} with a shift-invert spectral transformation, which in turn used LU decomposition and MUMPS~\cite{MUMPS:1,MUMPS:2}.
The stationary magnetization values were calculated as
\begin{align}
  & m_{k,\text{ss}} =\frac{2}{N} \sum_{J} \llangle J_k|\rho_J \rrangle \,,
  \end{align}
with the vector $\llangle J_k|$ defined as in Appendix~\ref{app:masteq-lmg}.

\begin{widetext}
\section{Derivation of auxiliary Hamiltonians for the CRF model} \label{app:auxhamcrf}

\subsection{Derivation for NO representation} \label{app:auxhamcrf-no}
Here we discuss the methods for deriving the auxiliary Hamiltonians for the CRF model. First, for the NO representation, we introduce the unnormalized distribution
\begin{align}
P_\text{NO} (v,v^*,z) \equiv\frac{1}{2 \pi^3} \int d^2 \xi d \eta \Tr[\hat{\rho} \hat{O}_\text{NO}] e^{-i\xi^* v^*} e^{-i\eta z} e^{-i\xi v} \,,
\end{align}
which is related to the normalized distribution $p_{\text{NO}}$ defined in Eq.~\eqref{eq:nodistribution-normalized} as
\begin{align}
p_\text{NO}(\boldsymbol{m})=N^3 P_\text{NO} (Nm_-,Nm_+,Nm_z) \,.
\end{align}
The distribution $P_\text{NO}$ evolves according to the equation of motion
\begin{align}
\partial_t P_\text{NO}(v,v^*,z)=\bar{L}_\text{NO}\left(v,v^*,z,\frac{\partial}{\partial v},\frac{\partial}{\partial v^*},\frac{\partial}{\partial z}\right) P_\text{NO}(v,v^*,z) \,.
\end{align}
The differential operator $\bar{L}_\text{NO}=\bar{L}_\text{NO}\left(v,v^*,z,\frac{\partial}{\partial v},\frac{\partial}{\partial v^*},\frac{\partial}{\partial z}\right)$ can be decomposed as
\begin{align}
\bar{L}_\text{NO}=\bar{L}_\Omega+\bar{L}_\Gamma+\bar{L}_\gamma+\bar{L}_\kappa \,,
\end{align}
where the successive terms on the right-hand side correspond to unitary dynamics, collective dissipation, local dissipation and local dephasing, respectively. The operator $\bar{L}_\text{NO}$ can be derived using a procedure presented in Ref.~\cite{haken1967quantum}; see also its textbook description in Chapter 6 of Ref.~\cite{carmichael1999statistical1}. Here we provide a brief outline of this framework. To obtain terms $\bar{L}_\mu$ ($\mu \in \{\Omega,\Gamma,\gamma,\kappa\}$) one first derives differential operators $M_\mu$ describing the corresponding evolution of the operator $\hat{O}_\text{NO}$ in the Heisenberg picture:
\begin{align} \label{eq:mopdefint}
\bar{L}_\mu P_\text{NO}(v,v^*,z)=\frac{1}{2 \pi^3} \int d^2 \xi d \eta \Tr[\hat{\rho} M_\mu \hat{O}_\text{NO}] e^{-i\xi^* v^*} e^{-i\eta z} e^{-i\xi v} \,,
\end{align}
with
\begin{align} \label{eq:mopdef}
\frac{i \Omega}{2}[\hat{S}_x,\hat{O}_\text{NO}] = M_\Omega \hat{O}_\text{NO}  \,, \quad \frac{\Gamma}{N} \mathcal{D}^\dagger [\hat{S}_-] \hat{O}_\text{NO} =M_\Gamma \hat{O}_\text{NO} \,, \quad \gamma \sum_{i=1}^N \mathcal{D^\dagger}[\sigma_-^i] \hat{O}_\text{NO}=M_\gamma \hat{O}_\text{NO} \,, \quad \frac{\kappa}{2} \sum_{i=1}^N \mathcal{D^\dagger}[\sigma_z^i] \hat{O}_\text{NO}=M_\kappa \hat{O}_\text{NO} \,,
\end{align}
where, recall, $\mathcal{D}^\dagger[\hat{A}] (\bullet) \equiv \hat{A}^\dagger (\bullet) \hat{A}-\{ \hat{A}^\dagger \hat{A},(\bullet) \}/2$. Here and from hereon, to keep our notation consistent with Ref.~\cite{carmichael1999statistical1}, we use a different convention to define collective spin operators, $\hat{S}_{x,y,z} \equiv 2 \hat{J}_{x,y,z}$ and $\hat{S}_{\pm} \equiv \hat{J}_\pm$.  The operators $M_\Omega$ and $M_\Gamma$ can be derived using the operator correspondence rules
\begin{align} \nonumber
&\hat{S}_+ \hat{O}_\text{NO} =\frac{\partial}{\partial (i \xi^*)} \hat{O}_\text{NO} \,, \, \hat{S}_- \hat{O}_\text{NO}=\left[e^{2i\eta} \frac{\partial}{\partial (i \xi)}-i \xi^* \frac{\partial}{\partial (i \eta)}-(i\xi^*)^2 \frac{\partial}{\partial (i \xi^*)} \right] \hat{O}_\text{NO} \,, \,  \hat{S}_z \hat{O}_\text{NO}=\left[\frac{\partial}{\partial(i \eta)}+2(i \xi^*) \frac{\partial}{\partial(i \xi^*)}\right] \hat{O}_\text{NO} \,,\\
&\hat{O}_\text{NO} \hat{S}_-=\frac{\partial}{\partial (i\xi)} \hat{O}_\text{NO} \,,  \,
\hat{O}_\text{NO} \hat{S}_+=\left[e^{2i\eta} \frac{\partial}{\partial (i \xi^*)}-i \xi \frac{\partial}{\partial (i \eta)}-(i\xi)^2 \frac{\partial}{\partial (i \xi)} \right] \hat{O}_\text{NO} \,, \, \hat{O}_\text{NO} \hat{S}_z =\left[\frac{\partial}{\partial(i \eta)}+2(i \xi) \frac{\partial}{\partial(i \xi)}\right] \hat{O}_\text{NO} \,,
\end{align}
with $\hat{S}_x=\hat{S}_++\hat{S}_-$. This yields
\begin{subequations}
\begin{align}
M_\Omega&=\frac{i \Omega}{2} \left[(e^{2i\eta} -1) \left(\frac{\partial}{\partial (i \xi)}-\frac{\partial}{\partial (i \xi^*)} \right)+(i\xi-i \xi^*) \frac{\partial}{\partial (i \eta)}-(i\xi^*)^2 \frac{\partial}{\partial (i \xi^*)}+(i\xi)^2 \frac{\partial}{\partial (i \xi)} \right] \,, \\
M_\Gamma &=\frac{\Gamma}{N} \left[(1-e^{2i\eta}) \frac{\partial^2}{\partial (i \xi) \partial (i \xi^*)}+\frac{i}{2} \xi^* \frac{\partial}{\partial (i \eta) \partial(i \xi^*)}+\frac{1}{2} (i\xi^*)^2 \frac{\partial^2}{\partial (i \xi^*)^2}+\frac{i}{2} \xi \frac{\partial}{\partial (i \eta) \partial(i \xi)}+\frac{1}{2} (i\xi)^2 \frac{\partial^2}{\partial (i \xi)^2} \right] \,.
\end{align}
\end{subequations}
To derive $M_\gamma$ one employs the expansion $\hat{O}_\text{NO}=\bigotimes_{i=1}^N \hat{O}^i_N$ where $\hat{O}^i_N=e^{i \xi^* \sigma_+^i} e^{i \eta \sigma_z^i} e^{i \xi \sigma_-^i}$. Then
\begin{align} \nonumber
\mathcal{D}^\dagger [\sigma_-^i] \hat{O}_\text{NO}&=\left(\sigma_+^i \hat{O}_\text{NO}^i \sigma_-^i -\sigma_+^i \sigma_-^i \hat{O}_\text{NO}^i/2-\hat{O}_\text{NO}^i \sigma_+^i \sigma_-^i/2\right) \bigotimes_{j \neq i} \hat{O}_\text{NO}^j \\ &=\left(\sigma_+^i \hat{O}_\text{NO}^i \sigma_-^i -\sigma_z^i \hat{O}_\text{NO}^i/4-\hat{O}_\text{NO}^i \sigma_z^i/4-\hat{O}_\text{NO}^i/2 \right) \bigotimes_{j \neq i} \hat{O}_\text{NO}^j \,,
\end{align}
where we employed $\sigma_+^i \sigma_-^i=(\mathds{1}+\sigma_z^i)/2$. We further use the identity $\sigma_+^i e^{i \eta \sigma_z^i}=e^{-2 i \eta} e^{i \eta \sigma_z^i} \sigma_+^i$ to write
\begin{align} \nonumber
\sigma_+^i \hat{O}_\text{NO}^i \sigma_-^i &=\sigma_+^ie^{i \xi^* \sigma_+^i} e^{i \eta \sigma_z^i} e^{i \xi \sigma_-^i} \sigma_-^i=e^{-2i\eta} e^{i \xi^* \sigma_+^i} e^{i \eta \sigma_z^i} \sigma_+^i \sigma_-^i e^{i \xi \sigma_-^i}=\frac{e^{-2 i \eta}}{2} e^{i \xi^* \sigma_+^i} e^{i \eta \sigma_z^i} (\mathds{1}+\sigma_z^i) e^{i \xi \sigma_-^i} \\ \label{eq:expansionsponsm} &=\frac{e^{-2i\eta}}{2}\left(1+\frac{\partial}{\partial(i\eta)} \right) \hat{O}_\text{NO}^i\,.
\end{align}
Inserting the above results into~\eqref{eq:mopdef}, we obtain
\begin{align}
M_\gamma \hat{O}_\text{NO}=\gamma\sum_{i=1}^N \mathcal{D}^\dagger [\sigma_-^i] \hat{O}_\text{NO}=\gamma \left[\frac{e^{-2i\eta}}{2}\left(N+\frac{\partial}{\partial(i\eta)} \right) \hat{O}_\text{NO}-\hat{S}_z \hat{O}_\text{NO}/4-\hat{O}_\text{NO} \hat{S}_z/4-\hat{O}_\text{NO}/2 \right]
\end{align}
and thus
\begin{align}
M_\gamma=\frac{\gamma}{2} \left[  \left(e^{-2 i \eta}-1 \right)\left(N+\frac{\partial}{\partial(i\eta)} \right)-(i \xi) \frac{\partial}{\partial(i \xi)}-(i \xi^*) \frac{\partial}{\partial(i \xi^*)} \right] \,.
\end{align}
Finally, to derive $M_\kappa$ one uses
\begin{align}
\mathcal{D}^\dagger [\sigma_z^i] \hat{O}_\text{NO}=\left(\sigma_z^i \hat{O}_\text{NO}^i \sigma_z^i -\sigma_z^i \sigma_z^i \hat{O}_\text{NO}^i/2-\hat{O}_\text{NO}^i \sigma_z^i \sigma_z^i/2\right) \bigotimes_{j \neq i} \hat{O}_\text{NO}^j=\left(\sigma_z^i \hat{O}_\text{NO}^i \sigma_z^i -\hat{O}_\text{NO}^i \right) \bigotimes_{j \neq i} \hat{O}_\text{NO}^j \,,
\end{align}
where we employed $(\sigma_z^i)^2=\mathds{1}$. Using further $(\sigma_\pm^i)^2=0$, $e^{a \sigma_\pm^i}=\mathds{1}+a \sigma_\pm^i$ and $\sigma_z^i \sigma_\pm^i=-\sigma_\pm^i \sigma_z^i$ we obtain
\begin{align} \nonumber
&\sigma_z^i \hat{O}_\text{NO}^i \sigma_z^i=\sigma_z^i(\mathds{1}+i \xi^* \sigma_+^i) e^{i \eta \sigma_z^i} (\mathds{1}+i \xi \sigma_-^i) \sigma_z^i=(\mathds{1}-i \xi^* \sigma_+^i) \sigma_z^i e^{i \eta \sigma_z^i} \sigma_z^i (\mathds{1}-i \xi \sigma_-^i) \\& =(\mathds{1}-2i \xi^* \sigma_+^i) (\mathds{1}+i \xi^* \sigma_+^i) (\sigma_z^i)^2 e^{i \eta \sigma_z^i}  (\mathds{1}+i \xi \sigma_-^i) (\mathds{1}-2i \xi \sigma_-^i) = (\mathds{1}-2 i\xi^* \sigma_+^i)\hat{O}^i_\text{NO} (\mathds{1}-2 i\xi \sigma_-^i) \,.
\end{align}
Inserting the above results into Eq.~\eqref{eq:mopdef} we obtain
\begin{align}
M_\kappa \hat{O}_\text{NO}=\frac{\kappa}{2} \sum_{i=1}^N \mathcal{D}^\dagger [\sigma_z^i] \hat{O}_\text{NO}=\kappa\left[-i\xi^* \hat{S}_+ \hat{O}_\text{NO}-i \xi \hat{O}_\text{NO} \hat{S}_-+2 (i \xi) (i \xi^*) \sum_{i=1}^N \sigma_+^i \hat{O}_\text{NO}^i \sigma_-^i \bigotimes_{j \neq i} \hat{O}_\text{NO}^j\right] \,.
\end{align}
Using~\eqref{eq:expansionsponsm} we thus get
\begin{align}
M_\kappa=\kappa \left[  e^{-2 i \eta} (i \xi)(i \xi^*)\left(N+\frac{\partial}{\partial(i\eta)} \right)-(i \xi) \frac{\partial}{\partial(i \xi)}-(i \xi^*) \frac{\partial}{\partial(i \xi^*)} \right] \,.
\end{align}

The operators $\bar{L}_\mu$ can then be formally derived by integrating the right-hand side of Eq.~\eqref{eq:mopdefint}. In practice, they can be easily obtained from the operators $M_\mu$ by performing the substitution~\cite{carmichael1999statistical1}
\begin{align}
\frac{\partial}{\partial (i \xi)} \rightarrow v \,, \, \frac{\partial}{\partial (i\xi^*)} \rightarrow v^* \,,\, \frac{\partial}{\partial (i\eta)} \rightarrow z \,,\, i\xi \rightarrow -\frac{\partial}{\partial v} \,,\, i\xi^* \rightarrow -\frac{\partial}{\partial v^*} \,,\, i\eta \rightarrow -\frac{\partial}{\partial z} \,.
\end{align}
This yields
\begin{subequations}
\begin{align}
\bar{L}_\Omega&=\frac{i\Omega}{2} \left[\left(e^{-2 \frac{\partial}{\partial z}}-1\right) (v-v^*)+ \frac{\partial}{\partial v^*} z -\frac{\partial}{\partial v} z +\frac{\partial^2}{\partial v^2} v-\frac{\partial^2}{\partial v^{*2}} v^* \right] \,, \\
\bar{L}_\Gamma&=\frac{\Gamma}{2N} \left[2 \left(1-e^{-2\frac{\partial}{\partial z}}\right) v v^*- \left(\frac{\partial}{\partial v} v+\frac{\partial}{\partial v^*} v^* \right)z +\left(\frac{\partial^2}{\partial v^2} v^2+\frac{\partial^2}{\partial v^{*2}} v^{*2} \right)\right] \,,
\\ \bar{L}_\gamma&=\frac{\gamma}{2}  \left[\left(e^{2 \frac{\partial}{\partial z}}-1\right)(N+z)+\frac{\partial}{\partial v} v+\frac{\partial}{\partial v^*} v^* \right] \,, \\
\bar{L}_\kappa&=\kappa \left[ \frac{\partial}{\partial v} v+\frac{\partial}{\partial v^*}v^* + \frac{\partial^2}{\partial v \partial v^*} e^{2 \frac{\partial}{\partial z}} (N+z) \right] \,.
\end{align}
\end{subequations}
We note that the expressions from $\bar{L}_\Omega$, $\bar{L}_\Gamma$ and $\bar{L}_\gamma$ agree with those from Refs.~\cite{haken1967quantum,agarwal1980fokker,lee2013dissipative,carmichael1999statistical1}. The result for $\bar{L}_\kappa$ differs from that from Ref.~\cite{carmichael1999statistical1} by a factor 2, but we carefully verified it to be correct. The operator $L_\text{NO}(\boldsymbol{m},\nabla_{\boldsymbol{m}})$ can then be obtained from $\bar{L}_\text{NO}$ by performing the substitution $v \rightarrow N(m_x-im_y)/2$, $z \rightarrow N m_z$, $\partial_v \rightarrow N^{-1} (\partial_{m_x}+i \partial_{m_y})$ and $\partial_z \rightarrow N^{-1} \partial_{m_z}$. Using Eq.~\eqref{eq:auxhamdef} with $\boldsymbol{x}=\boldsymbol{m}=(m_x,m_y,m_z)$ and $V=N$, we obtain the Hamiltonian explicitly presented in Eq.~\eqref{eq:auxham-crf-no}.

\subsection{Derivation for GC representation} \label{app:auxhamcrf-gc}
The derivation for GC representation starts from considering the differential operator $L_\text{GC}(\boldsymbol{m},\nabla_{\boldsymbol{m}})$ that describes the evolution of the distribution $p_\text{GC}$. It can be obtained using the equality
\begin{align}
\partial_t \hat{\rho}=\int p_\text{GC}(\boldsymbol{m}) \partial_t \hat{g}(\boldsymbol{m}) d\boldsymbol{m} \,.
\end{align}
The equations of motion for $\hat{g}(\boldsymbol{m})$ can be obtained using the operator correspondence rules for $k \in \{x,y,z \}$
\begin{subequations}
\begin{align}
\hat{J}_k \hat{g}(\boldsymbol{m}) &= \frac{1}{2} \left[N m_k+\sum_{l} (\delta_{kl}-m_k m_l-i \sum_{n} \varepsilon_{kln} m_{n}) \partial_{m_l}  \right] \hat{g} (\boldsymbol{m}) \,, \\
\hat{g}(\boldsymbol{m}) \hat{J}_k &= \frac{1}{2}\left[N m_k+\sum_{l} (\delta_{kl}-m_k m_l+i \sum_{n} \varepsilon_{kln} m_{n}) \partial_{m_l}  \right] \hat{g} (\boldsymbol{m}) \,,
\end{align}
\end{subequations}
and
\begin{subequations}
    \begin{align}
\sum_i^N \mathcal{D} \left(\sigma_{\pm}^i \right) [\hat{g}(\boldsymbol{m})] &= -\frac{1}{2} \left[m_x \partial_{m_x} +m_y\partial_{m_y} \mp 2 (1 \mp m_z) \partial_{m_z} \right] \hat{g}(\boldsymbol{m}) \,, \\
\sum_i^N \mathcal{D} \left(\sigma_z^i \right) [\hat{g}(\boldsymbol{m})] &= -2\left[m_x \partial_{m_x} +m_y\partial_{m_y}\right] \hat{g}(\boldsymbol{m}) \,.
\end{align}
\end{subequations}
In this way we obtain 
\begin{align}
\partial_t \hat{g}(\boldsymbol{m}) = L_\text{GC}^\dagger(\boldsymbol{m},\nabla_{\boldsymbol{m}}) \hat{g}(\boldsymbol{m}) \,,
\end{align}
where $L_\text{GC}^\dagger(\boldsymbol{m},\nabla_{\boldsymbol{m}})$ is the adjoint of $L_\text{GC}(\boldsymbol{m},\nabla_{\boldsymbol{m}})$. It can be expressed as
\begin{align}
L_\text{GC}^\dagger(\boldsymbol{m},\nabla_{\boldsymbol{m}})=\sum_k h_k \partial_{m_k}+\frac{1}{N} \sum_{k,l} D_{kl} \partial_{m_k} \partial_{m_l}
\end{align}
with
\begin{align} \nonumber
&h_x=\frac{\Gamma}{2} m_x m_z-\frac{\gamma}{2} m_x-\kappa m_x \,, \quad h_y=-\Omega m_z+\frac{\Gamma}{2} m_y m_z-\frac{\gamma}{2} m_y-\kappa m_y \,, \quad h_z=\Omega m_y-\frac{\Gamma}{2}(m_x^2+m_y^2)-\gamma(1+m_z) \,, \\ \nonumber
&D_{xx}=\frac{\Gamma}{2}(1-m_x^2+m_z)m_z \,, \quad D_{yy}=\frac{\Gamma}{2}(1-m_y^2+m_z)m_z \,, \quad D_{zz}=\frac{\Gamma}{2}(m_x^2+m_y^2)(1+m_z) \,,
 \\ 
&D_{xy}=D_{yx}=-\frac{\Gamma}{2} m_x m_y m_z \,, \quad D_{xz}=D_{zx}=\frac{\Gamma}{4} [m_x^2+m_y^2-(1+m_z)^2]m_x \,, \quad D_{yz}=D_{zy}=\frac{\Gamma}{4} [m_x^2+m_y^2-(1+m_z)^2]m_y \,.
\end{align}
The operator $L_\text{GC}(\boldsymbol{m},\nabla_{\boldsymbol{m}})$ can then be obtained from $L_\text{GC}^\dagger(\boldsymbol{m},\nabla_{\boldsymbol{m}})$ by changing the signs of the derivatives $\partial_{m_k}$ and moving the variables $m_k$ to the right of the derivatives. Equivalently, the Hamiltonian $\mathcal{H}_\text{GC}$ can be obtained directly from $L_\text{GC}^\dagger(\boldsymbol{m},\nabla_{\boldsymbol{m}})$ as
\begin{align}
\mathcal{H}_\text{GC}(\boldsymbol{m},\boldsymbol{\pi})=\lim_{N \rightarrow \infty} N^{-1} L_\text{GC}^\dagger(\boldsymbol{m},N\boldsymbol{\pi}) \,.
\end{align}
This yields the Hamiltonian explicitly presented in Eq.~\eqref{eq:auxham-crf-gc}.

\subsection{Derivation for SW approach} \label{app:auxhamcrf-sw}
The SW approach can be applied to the CRF model with a weak permutation symmetry following the method from Ref.~\cite{gronchi1979fokker}. One first derives the differential operator $\bar{L}_\text{SW}$ describing the evolution of the unnormalized Wigner distribution
\begin{align}
P_\text{W} (v,v^*,z) \equiv\frac{1}{2 \pi^3} \int d^2 \xi d \eta \Tr \left[\hat{\rho}  e^{i (\xi^* \hat{S}_+ +\eta \hat{S}_z +\xi \hat{S}_-)} \right] e^{-i\xi^* v^*} e^{-i\eta z} e^{-i\xi v} \,,
\end{align}
which is truncated after the second derivative term; as in Appendix~\ref{app:auxhamcrf-no} we use operators $\hat{S}_{x,y,z} \equiv 2 \hat{J}_{x,y,z}$ and $\hat{S}_{\pm} \equiv \hat{J}_\pm$. To that end, one first derives the equations of motion for expectation values of symmetrically ordered moments of operators
\begin{subequations}
\begin{align}
d_t \langle \hat{S}_z \rangle=&-\frac{2 \Gamma}{N} \langle \hat{S}_+ \hat{S}_- \rangle_S - i \Omega \left(\langle \hat{S}_+ \rangle-\langle \hat{S}_- \rangle \right)-\left(\frac{\Gamma}{N}+\gamma \right) \langle \hat{S}_z\rangle-\gamma N \,, \\
d_t \langle \hat{S}_\pm \rangle=&\frac{\Gamma}{2N} \langle \hat{S}_\pm \hat{S}_z \rangle_S-\left(\frac{\Gamma}{2N} +\frac{\gamma}{2}+\kappa \right) \langle \hat{S}_\pm\rangle \mp \frac{i \Omega}{2} \langle \hat{S}_z \rangle \,,\\ \nonumber
d_t \langle \hat{S}_z^2 \rangle=& -\frac{4 \Gamma}{N} \langle \hat{S}_+ \hat{S}_z \hat{S}_- \rangle_S +\frac{4 \Gamma}{N} \langle \hat{S}_+ \hat{S}_- \rangle_S-2i \Omega \left(\langle \hat{S}_+ \hat{S}_z \rangle_S - \langle \hat{S}_- \hat{S}_z \rangle_S \right)-2 \left(\frac{\Gamma}{N}+\gamma \right)  \langle \hat{S}_z^2 \rangle  \\ &+ \left[ \frac{2 \Gamma}{3N}-2\gamma(N-1) \right] \langle \hat{S}_z \rangle+2 \gamma N \,, \\
d_t \langle \hat{S}_\pm^2 \rangle=&\frac{\Gamma}{N} \langle \hat{S}_\pm^2 \hat{S}_z \rangle_S-\left( \frac{\Gamma}{N}+\gamma  +2 \kappa \right) \langle \hat{S}_\pm^2 \rangle \mp i \Omega\langle \hat{S}_\pm \hat{S}_z \rangle_S \,, \\
d_t \langle \hat{S}_+ \hat{S}_- \rangle_S=&\frac{\Gamma}{N} \langle \hat{S}_+ \hat{S}_z \hat{S}_- \rangle_S-\left(\frac{\Gamma}{N} + \gamma +2 \kappa
\right) \langle \hat{S}_+ \hat{S}_- \rangle_S + \frac{i \Omega}{2} \left(\langle \hat{S}_+ \hat{S}_z \rangle_S - \langle \hat{S}_- \hat{S}_z \rangle_S \right) +\frac{\Gamma}{2N}\langle \hat{S}_z^2 \rangle-\frac{\Gamma}{6N}\langle \hat{S}_z \rangle + N \left(\frac{\gamma}{2}+\kappa \right) \,,\\ \nonumber
d_t \langle \hat{S}_\pm \hat{S}_z \rangle_S =& -\frac{2  \Gamma}{N} \langle \hat{S}_\pm^2 \hat{S}_\mp \rangle_S-\left( \frac{5\Gamma}{2N}+\frac{3 \gamma}{2} +\kappa \right)\langle \hat{S}_\pm \hat{S}_z \rangle_S+\frac{\Gamma}{2N} \langle \hat{S}_\pm \hat{S}_z^2 \rangle_S+ \left[ \frac{\Gamma}{2N}-\gamma(N-1) \right] \langle \hat{S}_\pm \rangle \\
&\mp \frac{i \Omega}{2} \left(2 \langle \hat{S}_\pm^2 \rangle+\langle \hat{S}_z^2 \rangle-2 \langle \hat{S}_+ \hat{S}_- \rangle_S \right) \,,
\end{align}
\end{subequations}
where $\langle \hat{A}_1\dots\hat{A}_k \rangle_S$ denotes the expectation value of the symmetrized (i.e., equally weighted over all permutations) product of the operators $\hat{A}_1\dots\hat{A}_k$. Then, if the expectation values obey the equations of motion
\begin{align}
d_t \langle \hat{S}_+^p \hat{S}_z^q \hat{S}_-^r \rangle=\sum_{p'q'r'} C^{pqr}_{p'q'r'} \langle \hat{S}_+^{p'} \hat{S}_z^{q'} \hat{S}_-^{r'} \rangle 
\end{align}
then the differential operator takes the form
\begin{align}
\bar{L}_\text{SW}=\sum_{pqr}  \frac{\partial^{p+q+r}}{\partial (v^*)^p \partial z^q \partial v^r} \sum_{p'q'r'} d^{pqr}_{p'q'r'} (v^*)^{p'} z^{q'} v^{r'} \,,
\end{align}
with the coefficients
\begin{subequations}
\begin{align}
d^{pqr}_{p'q'r'} &=-C^{pqr}_{p'q'r'} \quad \text{for} \quad p+q+r=1 \,, \\
d^{pqr}_{p'q'r'} &=\frac{1}{p!q!r!} C^{pqr}_{p'q'r'}-C^{(p-1)qr}_{(p'-1)q'r'}-C^{p(q-1)r}_{p'(q'-1)r'}-C^{pq(r-1)}_{p'q'(r'-1)} \quad \text{for} \quad p+q+r=2 \,.
\end{align}
\end{subequations}
Explicitly, one obtains
\begin{align} \nonumber
&\bar{L}_\text{SW}=\frac{i\Omega}{2} \left[\frac{\partial}{\partial v^*} z -\frac{\partial}{\partial v} z +2 \frac{\partial}{\partial z} (v^*-v)\right] +\frac{\Gamma}{2N} \left[4\frac{\partial}{\partial z}\left(1+\frac{\partial}{\partial z} \right)vv^*-\left(1+2\frac{\partial}{\partial z} \right) \left(\frac{\partial}{\partial v} v+\frac{\partial}{\partial v^*}v^* \right)z \right. \\ \nonumber &\left. +\left(1+\frac{\partial}{\partial z} \right) \left(\frac{\partial}{\partial v} v+\frac{\partial}{\partial v^*}v^* \right)  +\frac{\partial^2}{\partial v \partial {v^*} }\left(z^2-\frac{z}{3} \right)+2 \frac{\partial}{\partial z} \left(1+\frac{1}{3} \frac{\partial}{\partial z} \right) z\right] \\ &+\frac{\gamma}{2} \left[2\frac{\partial}{\partial z}\left(1+\frac{\partial}{\partial z} \right)(N+z)+\left(1+2\frac{\partial}{\partial z} \right) \left(\frac{\partial}{\partial v} v+\frac{\partial}{\partial v^*}v^* \right)+N \frac{\partial^2}{\partial v \partial v^*}\right]
+\kappa \left( \frac{\partial}{\partial v} v+\frac{\partial}{\partial v^*}v^* + N \frac{\partial^2}{\partial v \partial v^*} \right) \,. \end{align}
The corresponding operator $L_\text{SW}$ for the normalized distribution $
p_\text{W}(\boldsymbol{m})=N^3 P_\text{W} (Nm_-,Nm_+,Nm_z) $ is then obtained as for NO representation,  by performing the substitution $v \rightarrow N(m_x-im_y)/2$, $z \rightarrow N m_z$, $\partial_v \rightarrow N^{-1} (\partial_{m_x}+i \partial_{m_y})$ and $\partial_z \rightarrow N^{-1} \partial_{m_z}$. Using Eq.~\eqref{eq:auxhamdef}  with $\boldsymbol{x}=\boldsymbol{m}=(m_x,m_y,m_z)$ and $V=N$, we obtain the Hamiltonian explicitly presented in Eq.~\eqref{eq:auxham-crf-sw}.

\subsection{Explicit forms of the auxiliary Hamiltonians} \label{app:auxhamcrf-res}
As a result of these derivations, we obtain the explicit forms of the auxiliary Hamiltonians for the CRF model,
\begin{subequations}
\begin{align} \nonumber
\mathcal{H}_\text{NO}(\boldsymbol{m},\boldsymbol{\pi})=&\frac{\Omega}{2} \left[\left(e^{2 \pi_z}-1 \right) m_y-2 m_z \pi_y -2 m_x \pi_x \pi_y +m_y(\pi_x^2-\pi_y^2)\right]  \\ \nonumber
&+\Gamma\left[ \left(1-e^{2\pi_z} \right) (m_x^2+m_y^2)/4+\left(m_x \pi_x+m_y \pi_y\right) m_z/2 +(m_x^2-m_y^2)(\pi_x^2-\pi_y^2 )/4+m_x m_y \pi_x \pi_y \right]
\\ \label{eq:auxham-crf-no}
&+\frac{\gamma}{2}  \left[(1+m_z) \left(e^{-2 \pi_z}-1\right)-m_x \pi_x-m_y \pi_y \right]+\kappa \left[(1+m_z) (\pi_x^2+\pi_y^2) e^{-2 \pi_z}-m_x \pi_x-m_y \pi_y \right] \,, \\ \nonumber
\mathcal{H}_\text{GC}(\boldsymbol{m},\boldsymbol{\pi})=&\Omega(m_y \pi_z-m_z \pi_y) +\frac{\Gamma}{2} \left\{(m_x^2+m_y^2)[(1+m_z)\pi_z^2-\pi_z] +[m_x^2+m_y^2-(1+m_z)^2](m_x \pi_x+m_y \pi_y) \pi_z  \right. \\ \nonumber
&\left. -2m_x m_y m_z \pi_x \pi_y +m_z [m_x \pi_x+m_y \pi_y +\pi_x^2(1-m_x^2+m_z) +\pi_y^2(1-m_y^2+m_z)]\right\} \\ \label{eq:auxham-crf-gc} &-\frac{\gamma}{2} \left[m_x \pi_x+m_y \pi_y+2(1+m_z)\pi_z \right] -\kappa(m_x \pi_x+m_y \pi_y) \,, \\ \nonumber
\mathcal{H}_\text{SW}(\boldsymbol{m},\boldsymbol{\pi})=&\Omega(m_y \pi_z-m_z \pi_y)   +\frac{\Gamma}{2}\left[ (\pi_z^2-\pi_z) (m_x^2+m_y^2)+(1-2 \pi_z)\left(m_x \pi_x+m_y \pi_y\right) m_z +m_z^2(\pi_x^2+\pi_y^2) \right]
\\ \label{eq:auxham-crf-sw}
&+\frac{\gamma}{2} \left[2(\pi_z^2-\pi_z)(1+m_z)-(1-2 \pi_z)(m_x \pi_x+m_y \pi_y)+\pi_x^2+\pi_y^2 \right] +\kappa \left( \pi_x^2+\pi_y^2 -m_x \pi_x-m_y \pi_y \right) \,.
\end{align}
\end{subequations}

\end{widetext}

\bibliography{bibliography}

\end{document}